\documentclass{llncs}

\usepackage{color}

\usepackage{times}
\usepackage{helvet}
\usepackage{courier}\title{Title}

\usepackage{verbatim}
\usepackage{mathptmx}
\usepackage{amssymb}
\usepackage{graphicx}
\usepackage{url}

\newcommand{\bott}{{\bot}}
\newcommand{\df}{\mathit{d}}
\newcommand{\rra}{\rightarrow}
\newcommand{\lra}{\leftarrow}
\newcommand{\vph}{\varphi}
\newcommand{\tr}{\mathit{true}}
\newcommand{\fa}{\mathit{false}}
\newcommand{\un}{unknown}
\newcommand{\nop}[1]{}
\newcommand{\I}{\mbox{$\cal I$}}
\newcommand{\mc}[1]{\marginpar{\scriptsize{#1}}}
\newcommand{\R}{{\cal R}}

\newcommand{\Dom}{\mathit{Dom}}
\newcommand{\LL}{\mathcal{L}}       % generic program
\newcommand{\At}{\mathit{At}}
\newcommand{\upar}{\Uparrow}
\newcommand{\doar}{\Downarrow}

\newcommand{\W}{\mbox{$\cal W$}}
\newcommand{\st}{\;|\,}

\newcommand{\datn}{\mbox{Datalog$^{\neg}$}}

\newcommand{\D}{\mbox{$\cal D$}}
\newcommand{\IC}{\eta}
\def\<{\mbox{$\langle$}}
\def\>{\mbox{$\rangle$}}
\newcommand{\grnd}{\mathit{ground}}

\newcommand{\U}{{\cal U}}
\newcommand{\J}{\mbox{$\cal J$}}

\newcommand{\RS}{\mbox{$\cal S$}}

\begin{document}

\title{The View-Update Problem for Indefinite Databases}

 \author {Luciano Caroprese\inst{1}, Irina Trubitsyna\inst{1}
          \and
          Miros{\l}aw Truszczy{\'n}ski\inst{2}
          \and
          Ester Zumpano\inst{1}
    }
\institute{DEIS, Universit\`a della Calabria, 87030 Rende, Italy
          \email{caroprese|irina|zumpano@deis.unical.it}
          \and
            Department of Computer Science, University of Kentucky,
         Lexington, KY~40506, USA
       \email{mirek@cs.uky.edu}
}
% \author {LUCIANO CAROPRESE, IRINA TRUBITSYNA\inst{1}
%          DEIS, Universit\`a della Calabria, 87030 Rende, Italy \\
%          \email{caroprese|irina@deis.unical.it}
%          \and
%          MIROS{\L}AW TRUSZCZY{\'N}SKI\\
%          Department of Computer Science, University of Kentucky,
%          Lexington, KY~40506, USA\\
%          \email{mirek@cs.uky.edu}
%          \and
%          ESTER ZUMPANO\\
%          DEIS, Universit\`a della Calabria, 87030 Rende, Italy \\
%          \email{zumpano@deis.unical.it}
%    }

%%\volume{\textbf{10} (3):}
%%\jdate{March 2002}
%\setcounter{page}{1}
%%\pubyear{2002}

\maketitle

%\label{firstpage}

\begin{abstract}
This paper introduces and studies a declarative framework for
updating views over \emph{indefinite data\-bases}. An indefinite
database is a database with null values that are represented,
following the standard database approach, by a single \emph{null}
constant. The paper formalizes views over such databases as
indefinite \emph{deductive} databases, and defines for them several
classes of database repairs that realize view-update requests. Most
notable is the class of \emph{constrained repairs}. Constrained
repairs change the database ``minimally'' and avoid making arbitrary
commitments. They narrow down the space of alternative ways to
fulfill the view-update request to those that are grounded, in a
certain strong sense, in the database, the view and the view-update
request.
\end{abstract}
%\begin{keywords}
%View Updating, Deductive Databases, Indefinite Databases, Semantic Foundations,
%Knowledge Representation, Abduction.
%\end{keywords}

\vspace*{-0.2in}
\section{Introduction}

A typical database system is large and complex. Users and
applications rarely can access the entire system directly.
Instead, it is more common that access is granted in terms of a
\emph{view}, a \emph{virtual} database consisting of relations
defined by a query to the \emph{stored and maintained}
database. Querying a view does not present a conceptual
problem. In contrast, another key task, \emph{view updating},
poses major challenges.

\begin{example}\label{example-der-pred}
Let $D=\{q(a,b)\}$ be a database over relation symbols $q$ and
$r$, where the relation $r$ has arity three and is currently empty.
Let us consider the view over $D$ given by the Datalog
program $P=\{p(X) \leftarrow q(X,Y),r(X,Y,Z)\}$. That
view consists of a single unary relation $p$. Given the present
state of $D$, the view is empty.

%Let us consider a situation when the user requests that $p(a)$ holds in
%the view (as it is now, it does not).
To satisfy the request that $p(a)$ holds in the view (as it is now, it
does not), one needs to update the database $D$. %so that after the
%update $p(a)$ holds in the
Such update consists of executing update actions that specify facts
to insert to and to delete from $D$. These update actions (in a
simplified setting that we consider for now) are ``signed'' facts
$+F$ and $-G$, where $+F$ stands for ``insert
F'' and $-G$ stands for ``delete G.'' In our case, the set of update
actions $\{-q(a,b), +q(a,a), +r(a,a,a)\}$ is a correct update to $D$.
Executing it on $D$ results in the database $D'=\{q(a,a), r(a,a,a)\}$,
which has the desired property that $p(a)$ holds in the view determined
by $P$. There are
also other ways to satisfy the user's request, for instance: $\{+r(a,b,a)\}$
and $\{+q(c,d), +r(c,d,d)\}$, where $c$ and $d$ are any elements of the
domain of the database.
\hfill $\Box$
\end{example}

As this example suggests, view updating consists of translating a
\emph{view-update request}, that is, an update request against the view,
into an \emph{update}, a set of update actions against the stored
(extensional)
database. The example highlights the basic problem of view updating. It
may be (in fact, it is common), that a view-update request can
be fulfilled by any of a large number of database updates. One
of them has to be committed to and executed. Thus, developing methods to
automate the selection process or, at least, aid the user in making the
choice is essential and has been one of the central problems of view
updating \cite{kakas90,cons95,TO95,MT03}.  That problem is also the focus
of our paper.\footnote{In some cases no update satisfies the view-update
request. The question whether the lack of an appropriate update is caused
by errors in the design of the view, in the extensional database, or in
the view-update request is interesting and deserves attention, but is
outside the scope of the present work.}

To restrict the space of possible database updates to select from, it is
common to consider only those that accomplish the view-update request and
are in some sense minimal.
That reduces the space of updates. For instance, the update
$\{-q(a,b), +q(a,a),$
$+r(a,a,a)\}$ in Example \ref{example-der-pred} is not minimal.
We can remove $-q(a,b)$ from it and what remains is still an update that
once executed ensures that $p(a)$ holds in the view.
Minimal updates
fulfilling the view-update request are commonly called \emph{repairs}.

Even imposing minimality may still leave a large number of
candidate repairs. In Example \ref{example-der-pred}, updates $\{+r(a,b,
\psi)\}$, %and $\{+q(a,\varphi),+r(a,\varphi,\psi)\}$,
where %$\varphi$ and
$\psi$ is any domain element, %and $\varphi\not=b$,
are repairs, and there are still more. As long as we
insist on the completeness of the repaired database, there is little one
can do at this point but ask the user to \emph{select} one.

The situation changes if we are willing to allow indefiniteness
(incomplete information) in databases. Given the ``regular'' structure
of the family of repairs above, one could represent it by a
single \emph{indefinite} repair, $\{+r(a,b,\psi)\}$
%and $\{+q(a,\varphi), +r(a,\varphi,\psi)\}$, respectively,
regarding %$\varphi$ and
$\psi$ as a distinct \emph{null} value standing for
some unspecified domain elements. The choice facing the user
substantially simplifies as possibly infinitely many repairs is reduced
to only one.

This approach was studied by Farr\'{e} et al.
%\citeyear{Ten*2003},
\cite{Ten*2003},
where the terminology of Skolem constants rather than null
values was used. However, while seemingly offering a plausible
solution to the problem of multiple repairs, the approach
suffers from two drawbacks. First, the use of multiple Skolem
constants (essentially, multiple null values) violates the SQL
standard \cite{TurkerG01}. Second, the approach assumes that
the initial database is complete (does not contain null
values). Thus, while the approach might be applied once,
iterating it fails as soon as an indefinite database is
produced, since no guidance on how to proceed in such case is
given.

We consider the view-update problem in the setting when \emph{both} the
original and the updated databases are indefinite (may contain
occurrences of null values). %In this way, we address the
%shortcoming of the approach by Farr\'{e} et al.
To be compliant with the SQL standard, we allow a single null value only.
We denote it by $\perp$ and interpret it as expressing the existence of
unknown values for each attribute it is used for \cite{Libkin95}. Typically,
a database is represented as a set of facts. We propose to represent
indefinite databases by \emph{two} sets of facts that we interpret by
means of a two-level closed-world assumption tailored to the case of
indefinite information. We interpret indefinite databases in terms of
their \emph{possible worlds}. We extend the possible-world semantics to
the setting of views, which we formalize in terms of \emph{indefinite
deductive databases}, and apply the resulting formalism to state and
study the view-updating problem.

We then turn attention to the problem of multiple repairs. In general,
using null values to encode multiple repairs is still not enough to
eliminate them entirely (cf. our discussion above), and some level of
the user's involvement may be necessary. Therefore, it is important to
identify principled ways to narrow down the space of repairs for the user
to consider. As the setting we consider here, when  both the original and
the updated databases may contain null values has not been considered
before, we propose a concept of minimality tailored precisely to that
situation. The primary concern is to minimize the set of new constants
introduced by an update. The secondary concern is to minimize the degree
of the semantic change. The resulting notion of minimality yields the
notion of a \emph{repair}.

Our concept of minimality leads us also to the concept of a \emph{relevant}
repair,
an update that introduces no new constants and minimizes the degree of
change. In Example \ref{example-der-pred}, $\{+r(a,b,\bot)\}$,
$\{+r(a,a,\bot),+q(a,a)\}$ and $\{+r(a,c,\bot), +q(a,c)\}$, where $c$ is
an element of the database domain other than $a$ and $b$, are all repairs.
%(we formally show these claims later).
The first two are obviously relevant,
the third one is not.

Some occurrences of non-nullary constants in a relevant repair may still
be ``ungrounded'' or ``arbitrary,'' that is, replacing them with another
constant results in a repair. For instance, replacing in the relevant
repair $\{+r(a,a,\bot),+q(a,a)\}$ the second and the forth
occurrences of $a$ with a fresh constant $c$ yields
$\{+r(a,c,\bot),+q(a,c)\}$, an update that is a repair. Intuitively,
``arbitrary'' occurrences of constants, being replaceable, are not
forced by the information present in the view (deductive database) and
in the view-update request. By restricting relevant repairs to those
\emph{without} arbitrary occurrences of constants we arrive at the class
of \emph{constrained} repairs. % that seem to be especially well suited as
%candidate repairs to enforce view-update requests.
In the view-update problem considered in Example
\ref{example-der-pred}, there is only \emph{one} constrained repair,
$\{+r(a,b,\bot)\}$.

Finally, we study the complexity of the problems of the existence of repairs,
relevant
repairs and constrained repairs. We obtain precise results for the first two
classes and an upper bound on the complexity for the last class. %Establishing
%tight results for the case of constrained repairs seems to be a challenging
%problem.

To summarize, our main contributions are as follows. We propose a
two-set representation of indefinite database that is more expressive than
the standard one. We define the semantics and the operation of updating
indefinite databases (Section \ref{Preliminaries}). We define views
over indefinite databases (indefinite deductive databases), and generalize
the semantics of indefinite databases to views (Section \ref{idd}). We
state and study the view-update problem in the general setting when the
initial and the repaired databases are indefinite. We propose a notion of
minimality of an update and use it to define the concept of a repair. We
address the problem of multiple repairs by defining relevant and constrained
repairs (Section \ref{vus}). We study the complexity of
problems of existence of repairs, relevant repairs and constrained
repairs (Section \ref{cs}). The proofs are in the appendix.

\vspace*{-0.1in}
\section{Indefinite Databases}
\label{Preliminaries}

We consider a finite set $\Pi$ of relation symbols and a set $\Dom$
of constants that includes a designated element $\bot$, called the
\emph{null value}. We define $\Dom_d=Dom\setminus\{\bot\}$.
Normally, we assume that $\Dom$ is an infinite
countable set. However, for the sake of simplicity, in several of the
examples the set $\Dom$ is finite.

Some predicates in $\Pi$ are designated as \emph{base} (or
\emph{extensional}) predicates and all the remaining ones are
understood as \emph{derived} (or \emph{intensional}) predicates.
A \emph{term} is a constant from $\Dom$ or a variable. An \emph{atom} is
an expression of the form $p(t_1,\ldots,t_k)$, where $p\in \Pi$ is a
\emph{predicate symbol} of arity $k$ and $t_i$'s are terms.
An atom is \emph{ground} if it does not contain variables. We refer to
ground atoms as \emph{facts}. We denote the set of all facts by $\At$.
We call facts defined in terms
of base and derived predicates \emph{base facts} and
\emph{derived facts} respectively.
%We write $\At^b$ and $\At^d$ for the sets of facts defined in terms
%of base and derived predicates, respectively. We call them \emph{base
%facts} and \emph{derived facts} respectively.
A fact is \emph{definite} if it does not
contain occurrences of $\bot$. Otherwise, it is \emph{indefinite}.
Given a set $S$ of atoms, we define $Dom(S)$ (resp. $Dom_d(S)$) as the set 
of constants in $Dom$ (resp. $Dom_d$) occurring in $S$.
For every two tuples of terms $t=(t_1,\ldots,t_k)$ and $t'=(t'_1,\ldots,
t'_k)$ and every $k$-ary predicate symbol $p\in\Pi$, we write $t\preceq
t'$ and  $p(t) \preceq p(t')$ if for every $i$, $1\leq i\leq k$, $t_i =
t'_i$ or $t_i =\perp$. We say in such case that $t'$ and $p(t')$ are
\emph{at least as informative} as $t$ and $p(t)$, respectively. If, in
addition, $t\not= t'$, we write $t\prec t'$ and $p(t)\prec p(t')$, and
say that $t'$ and $p(t')$ are \emph{more informative} than $t$ and
$p(t)$. Sometimes, we say ``at most as informative'' and ``less
informative,'' with the obvious understanding of the intended meaning.
We also define $t$ and $t'$ (respectively, $p(t)$ and $p(t')$) to be
\emph{compatible}, denoted by $t\approx t'$ (respectively, $p(t)\approx
p(t')$), if for some $k$-tuple $s$ of terms, $t \preceq s$ and $t'
\preceq s$. Finally, for a set $D\subseteq\At$, we define
\vspace*{-2mm}
\smallskip
\[
\begin{array}{llll}
\hspace{-8mm}D^\doar&\hspace{-2mm}=\hspace{0.5mm}\{a \st \mbox{there is $b\in D$ s.t. $a\preceq b$}\} &
\mbox{\quad}D^\upar&\hspace{-2mm}=\hspace{0.5mm}\{a\st \mbox{there is $b\in D$ s.t. $b\preceq a$}\}\\
\hspace{-8mm}D^\approx&\hspace{-2mm}=\hspace{0.5mm}\{a\st \mbox{there is $b\in D$ s.t. $b\approx a$}\}   &
\mbox{\quad}D^\sim&\hspace{-2mm}=\hspace{0.5mm}D^\approx\setminus D^\doar.%\\
%\hspace{-8mm}\left\lfloor D\right\rfloor&\hspace{-2mm}=\hspace{0.5mm}\{a\in D\st \mbox{for no $b\in D$,
%$b\prec a$}\}  &
%\mbox{\quad}\left\lceil  D\right\rceil&\hspace{-2mm}=\hspace{0.5mm}\{a\in D\st \mbox{for no $b\in  D$,
%$a\prec b$}\}.
\end{array}
\]
To illustrate, let $q$ be a binary relation symbol and $\Dom= \{\bot,1,2\}$. Then: %We then have

\noindent
\mbox{\quad}$\{q(1,\bot)\}^\doar=\{q(1,\bot), q(\bot,\bot)\}$ \hspace{12mm}
\mbox{\quad}$\{q(1,\bot)\}^\upar=\{q(1,\bot), q(1,1),q(1,2)\}$\\
\mbox{\quad}$\{q(1,\bot)\}^\approx=\{q(1,\bot),q(1,1),q(1,2),q(\bot,1),q(\bot,2),q(\bot,\bot)\}$\\
\mbox{\quad}$\{q(1,\bot)\}^\sim=\{q(1,1),q(1,2),q(\bot,1),q(\bot,2)\}$.
%\mbox{\quad}$(\{q(1,\bot)\}^\upar)^\doar=\{q(1,\bot),q(1,1),q(1,2),q(\bot,1),q(\bot,2),q(\bot,\bot)\}$.

\noindent
We note also note that $D^\approx = ( D^\upar)^\doar$.

In the most common case, %(extensional)
databases are finite subsets of
$\At$ that contain \emph{definite} facts only. The semantics of such
databases is given by the \emph{closed-world assumption} or CWA \cite{re78}:
a definite fact $q$ is \emph{true} in a database $ D$ if $q\in D$.
Otherwise, $q$ is \emph{false} in $ D$. We are interested in
databases that may contain indefinite facts, too.
Generalizing, we will \emph{for now} assume that an
indefinite database is a finite set of possibly indefinite atoms.
The key question is that of the semantics of indefinite databases.

Let $ D$ be an indefinite database. Clearly, all facts in $ D$ are true
in $ D$. In addition, any fact that is less informative than a fact in
$ D$ is also true in $ D$. Indeed, each such fact represents an
existential statement, whose truth is established by the presence of a
more informative fact in $ D$ (for instance, the meaning of $p(\bot)$ is
that there is an element $c$ in the domain of the database such that
$p(c)$ holds; if $p(1) \in D$, that statement is true).
Summarizing, every fact in $ D^\doar$ is true in $ D$.

By CWA adapted for the setting of indefinite
databases \cite{Libkin95}, facts that are not in $ D^\doar$ are
\emph{not} true in $ D$, as $ D$ contains no evidence to support
their truth. Those facts among them that are compatible with facts in
$ D$ (in our notation, facts in $ D^\sim$), might actually be true, but
the database just does not know that. Of course, they may also be false,
the database does not exclude that possibility either. They are regarded
as \emph{unknown}. By CWA again, the facts that
are not compatible with any fact in $ D$ are false in $ D$, as $ D$
provides no explicit evidence otherwise.

The simple notion of an indefinite database, while intuitive and having
a clear semantics, has a drawback. It has a limited expressive power.
For instance, there is no database $ D$ to represent our knowledge that
$p(1)$ is false and that there is some definite $c$ such that $p(c)$
holds (clearly, this $c$ is not $1$). To handle such cases, we
introduce a more general concept of an indefinite database, still using
CWA to specify its meaning.

\begin{definition}
An \emph{indefinite database} (a \emph{database}, for short) is a pair
$\I=\<D,E\>$, where $D$ and $E$ are finite sets of (possibly indefinite)
facts.\hfill$\Box$
\end{definition}

The intended role of $D$ is to represent all facts that are true in the
database $\<D,E\>$, while $E$ is meant to represent \emph{exceptions},
those facts that normally would be unknown, but are in fact false (and
the database knows it). More formally, the semantics of indefinite databases
is presented in the following definition.

\begin{definition}
Let $\<D,E\>$ be a database and let $q\in\At$ be a fact. Then:
%\begin{enumerate}
%\item
(1) $q$ is \emph{true} in $\<D,E\>$, written $\<D,E\>\models q$,
if $q\in D^\doar$;
%\item
(2) $q$ is \emph{unknown} in $\<D,E\>$, if $q\in( D^\approx\setminus
 D^\doar)\setminus E^\upar$ ($= D^\sim\setminus E^\upar$);
%\item
(3) $q$ is \emph{false} in $\<D, E\>$, denoted $\<D, E\>\models \neg q$,
in all other cases, that is, if $q\notin  D^\approx$ or if
$q\in D^\sim\cap E^\upar$. \hfill$\Box$
%\end{enumerate}
\end{definition}

The use of $ E^\upar$ in the definition (items (2) and (3)) reflects
the property that if an atom $a$ is false then every atom $b$ at least
as informative as $a$ must be false, too.

We denote the sets of all
facts that are true, unknown and false in a database $\I=\<D, E\>$ by
$\I_t$, $\I_u$ and $\I_f$, respectively. Restating the definition we
have:
\[
\I_t= D^\doar,\quad \I_u= D^\sim \setminus E^\upar,\mbox{\ and}\quad\I_f=(\At
\setminus  D^\approx) \cup ( D^\sim\cap E^\upar)\mbox{.}
\]
%$\I_t= D^\doar$, $\I_u= D^\sim \setminus E^\upar$,
%$\I_f=(\At\setminus  D^\approx) \cup ( D^\sim\cap E^\upar)$.

\begin{example}
The knowledge that $p(c)$ holds for some constant $c$ and that
$p(1)$ is false can be captured by the database $\<\{p(\bot)\},
\{p(1)\}\>$. The database
$\<\{q(\bot,\bot),q(1,1)\},$ $\{q(1,\bot)\}\>$ specifies that the
atoms $q(1,1)$, $q(1,\bot)$, $q(\bot,1)$ and $q(\bot,\bot)$ are true,
that all definite atoms $q(1,d)$, with $d\not=1$, are false, and
that all other atoms $q(c,d)$ are unknown. While the fact that
$q(\bot,\bot)$ is true follows from the fact that $q(1,1)$ is true, the
presence of the former in the database is not redundant, that is, the database
without it would have a different meaning. Namely, $q(\bot,\bot)$ makes
all atoms $q(a,b)$ potentially unknown (with some of them true or false
due to other elements of the database).
\hfill$\Box$
\end{example}

%Next, we define a \emph{possible world}, that is, a possible
%two-valued realization of a database.
%
\begin{definition}
\label{poss-world}
A set $W$ of \emph{definite} facts is a \emph{possible world} for a database
$\I=\<D, E\>$ if $\I_t\subseteq W^\doar$ ($W$ ``explains'' all facts that are
true in $\I$), $W\subseteq \I_t\cup \I_u$ (only definite facts that are
true or unknown appear in a possible world).
\hfill$\Box$
\end{definition}

%\noindent
%We observe that the problem of checking if
%an atom belongs to $D^\upar$, $D^\doar$ and $D^\approx$ can be solved in
%linear time. Thus, the problem of checking whether a set $W$ is a possible world of a database $\I$ can be solved in polynomial time in the size of $W$ and the database.

\noindent
Databases represent sets of possible worlds. For a database $\I$, we
write $\W(\I)$ to denote the family of all possible worlds for $\I$.
Due to the absence of indefinite facts in $W\in\W(\I)$, every fact in
$\At$ is either true (if it belongs to $W^\doar$) or false (otherwise) w.r.t. $W$.
Extending the notation we introduced earlier, for a possible world $W$
and $a\in\At$ we write $W\models a$ if $a\in W^\doar$ and $W\models \
\neg a$, otherwise.
The following proposition shows that the semantics of a database can be
stated in terms of its possible worlds.

\begin{proposition}
\label{prop:1}
Let $\I$ be a database and $q$ a fact. Then
$q\in\I_t$ if and only if $W\models q$, for every $W\in\W(\I)$, and
$q\in\I_f$ if and only if $W\models \neg q$, for every $W\in\W(\I)$.~\hfill$\Box$
\end{proposition}

%We now turn attention to the task of updating indefinite databases.
\emph{Updating} a database $\<D,E\>$ consists of executing on it \emph{update
actions}: inserting a base fact $a$ into $D$ or $E$, and deleting a base
fact $a$ from $D$ or $E$. We denote them by $+a^{D}$, $+a^{E}$, $-a^{D}$
and $-a^{E}$,
respectively. For a set ${U}$ of update actions, we define
${U}_{+}^D=\{a\st +a^{D}\in{U}\}$, ${U}_{+}^E=\{a\st +a^{E} \in{U}\}$,
${U}_{-}^D=\{a\st -a^{D}\in{U}\}$, and ${U}_{-}^E=\{a\st -a^{E}\in{U}\}$. To be
executable, a set ${U}$ of update actions must not contain \emph{contradictory}
update actions: $+a^{D}$ and $-a^{D}$, or $+a^{E}$ and $-a^{E}$. A
contradiction-free set ${U}$ of update actions is an \emph{update}. We denote
the set of all updates (in the fixed language we consider) by $\U$.

We now define the operation to update a database, that is, to apply an update
to it.

\begin{definition}
\label{defUpdate}
Let $\I=\<D, E\>$ be an indefinite database and ${U}$ an update.
We define $\I \circ {U}$ as the database $\<D', E'\>$, where
$D'=(D\cup{U}_{+}^D) \setminus {U}_{-}^D$
and
$ E'= (E\cup{U}_{+}^E)\setminus {U}_{-}^E$.
\hfill$\Box$
\end{definition}

%By selecting an update appropriately, any indefinite database can be
%transformed into any other indefinite database.
%
%\begin{proposition}\label{prop:2}
%Let $\I$ and $\I'$ be databases. There is an update ${U}\in\U$ such that
%$\I\circ {U}=\I'$.~\hfill$\Box$
%\end{proposition}
%

\section{Indefinite Deductive Databases}
\label{idd}

\emph{Integrity constraints} (ICs, for short) are first-order sentences
in the language $\LL$ determined by the set of predicates $\Pi$ and by
the set $\Dom_d$ of definite constants. A database with integrity constraints
is a pair $\<\I,\eta\>$, where $\I$ is a database and $\eta$ is a set of
ICs. Possible worlds can be regarded as interpretations of the language
$\LL$ with $\Dom_d$ as their domain. This observation and Proposition
\ref{prop:1} suggest a definition of the semantics of databases with ICs.

\begin{definition}
Let $\<\I,\eta\>$ be a database with ICs. A possible world $W\in\W(\I)$
is a \emph{possible world for a database $\<\I,\eta\>$} if $W$ satisfies
every integrity constraint in $\eta$. We denote the set of possible
worlds of $\<\I,\eta\>$ by $\W(\I,\eta)$.  A database with ICs, $\<\I,
\IC\>$, is \emph{consistent} if $\W(\I,\eta)\not=\emptyset$. Otherwise,
$\<\I,\IC\>$ is \emph{inconsistent}.

A fact $q$ is true in $\<\I,\eta\>$ if $W\models q$, for every
$W\in\W(\I,\eta)$; $q$ is false in $\<\I,\eta\>$ if $W\models \neg q$,
for every $W\in\W(\I,\eta)$; otherwise, $q$ is unknown in $\<\I,\eta\>$.
\hfill$\Box$
\end{definition}

\begin{example}
\label{ex:cons}
Let us consider the database $\I=\<\{p(1), p(2), q(\bot)\},\emptyset\>$
(there are no exceptions) and
let $\IC=\{\forall X (q(X)\rightarrow p(X))\}$. Possible worlds of $\I$
include $\{p(1), p(2),$ $ q(1)\}$, $\{p(1),p(2),$ $q(2)\}$ and $\{p(1),p(2),
q(3)\}$. The first two satisfy the integrity constraint, the third one
does not. Thus, only the first two are possible worlds of $\<\I,\IC\>$.
Since $p(1)$ belongs to all possible worlds of $\<\I,\IC\>$, $p(1)$ is
true in $\<\I,\IC\>$. Further, $p(3)$ is false in every possible world of
$\I$ and so also in every possible world of $\<\I,\eta\>$. Thus, $p(3)$
is false in $\<\I,\eta\>$. Lastly, we note that $q(1)$ and $q(2)$ are unknown in
$\<\I,\eta\>$, while $q(3)$ is false (due to $p(3)$ being false and the integrity
constraint). %even though it
%is not true in $\I$. Its truth in $\<\I,\IC\>$ is enforced by the integrity
%constraints, which limit the family of possible worlds.
\hfill$\Box$
\end{example}

We note that the possible-world semantics can capture additional
information contained in integrity constraints. In Example
\ref{ex:cons}, the semantics derives that $q(3)$ is \emph{false} in
$\<\I,\eta\>$ even though this knowledge is not present in the database
$\I$.

The concepts of an update and of the operation to execute an update on
a database extend literally to the case of databases with ICs.

Following Ullman %\citeyear{Ull98}, \emph{views} are safe \datn\
\cite{Ull98}, \emph{views} are safe \datn\
programs. We use the standard terminology and talk about (\datn)
\emph{rules}, and \emph{bodies} and \emph{heads} of rules. A rule is
\emph{safe} if each variable occurring in the head or in a
negative literal in the body also occurs in a positive literal in
the body. A \datn\ program is safe if each rule is \emph{safe}.
We assume that views do not contain occurrences of $\bot$.
The semantics of \datn programs is given in terms of \emph{answer
sets} \cite{GelLif88,GelLif91}. A precise definition of that semantics
is immaterial to our study and so we do not provide the details.

\begin{definition}
An \emph{indefinite deductive database} (from now, simply, a deductive
database) is a tuple $\D=\<\I,\IC,{P}\>$, where
$\I$ is a database, $\eta$ is a set of integrity constraints, and
${P}$ is a safe \datn\ program (the specification of a \emph{view})
such
that no predicate occurring in the head of a rule in ${P}$ is a base
predicate. ~\hfill$\Box$
\end{definition}

Clearly, a deductive database with the empty view is a database with ICs,
and a deductive database with the empty view and no ICs is simply a
database.

\begin{definition}
\label{consistent-DD}
A deductive database $\D=\<\I,\IC,{P}\>$ is
\emph{consistent} if $\<\I,\IC\>$ is consistent and for every
possible world $W\in \W(\I,\IC)$, the program $W\cup{P}$ has answer
sets. We denote the family of all those answer sets by $\W(\D)$ or
$\W(\I,\IC,{P})$. We call elements of $\W(\D)$
\emph{possible worlds} of $\D$. ~\hfill$\Box$
\end{definition}

There is an alternative to our concept of consistency.
One could define a deductive database $\<\I,\IC,{P}\>$ as consistent
if for \emph{at
least
one} world $W\in \W(\I,\IC)$, the program $W\cup{P}$ has an answer set.
That concept of consistency would allow situations where for some possible
worlds of $\<\I,\IC\>$, one of which could be a description of the real
world, the view ${P}$ does not generate any meaningful virtual database.
Our concept of consistency is more robust. It guarantees that
the user can have a view of a database no matter how the real world looks
like, that is, which of the possible worlds describes it.

\begin{example}
\label{ex:cons2} Let $\D=\<\I,\IC,{P}\>$ be a deductive database, where
$\I=\<\{p(\bot)\},\emptyset\>$, $\IC= \emptyset$ and ${P}=\{t \leftarrow
p(1), p(2),not\; t\}$, for some derived ground atom $t$. Every non-empty
set $W \subseteq \{p(u)\st u\in \Dom_d\}$ is a possible world of $\<\I,
\IC\>$. In particular, the set $\{p(1),p(2)\}$ is a possible
world of $\<\I,\IC\>$. Since the program ${P}\cup\{p(1),p(2)\}$ has no
answer sets, $\D$ is inconsistent (according to our definition). If
$\D'=\<\I,\{p(1)\wedge p(2) \rightarrow \bott\},{P}\>$, then the integrity
constraint in $\D'$ eliminates the offending possible world and one can
check that for every possible world $W$ of $\<\I,\{p(1)\wedge p(2)
\rightarrow \bott\}\>$, ${P}\cup W$ has an answer set. Thus, $\D'$ is
consistent.\footnote{We point out that in the paper,
we overload the notation $\bot$.
We use it to denote both the single null value in the language and the
falsity symbol in the first-order language used for integrity constraints.
Since the meaning is always clear from the context, no ambiguity arises.}
\hfill$\Box$
\end{example}

The concept of an update extends in a natural way to
deductive databases. If $U$ is an update, and $\D= \<\I,\IC,{P}\>$, we
define the result of updating $\D$ by $U$ by $\D\circ U = \<\I\circ U, \IC,
{P}\>$.

Next, we define the semantics of a deductive database $\D= \<\I,\IC,{P}\>$,
again building on the characterization given by Proposition \ref{prop:1}.
%We consider \emph{consistent} databases only, assuming that only
%consistent databases are queried (the problem of \emph{repairing}
%inconsistent databases is of course of interest; it can be seen as a
%special view-update problem that we address in the next section).

\begin{definition}%\mc{change}
\label{def:ddb}
A fact $a\in\At$ is \emph{true} in a deductive database $\D=\<\I,\IC,{P}\>$,
denoted by $\D \models a$, if for every possible world $W\in\W(\D)$,
$W\models a$; $a$ is \emph{false} in $\D$, denoted by $\D \models \neg a$,
if for every $W\in\W(\D)$, $W\models\neg a$; $a$ is \emph{unknown} in $\D$,
otherwise. We denote the truth value of $a$ in $\D$ by $v_{\cal{D}}(a)$.
%
%We denote the sets of facts that are \emph{true}, \emph{false} and
%\emph{unknown} in $\D$ by $\D_t$, $\D_f$ and $\D_u$, respectively.
%For a fact $a$, we denote its truth value in $\D$ by $v_{\cal{D}}(a)$.
\hfill$\Box$
\end{definition}

\begin{example}\label{ex:ded-db}
Let $\Dom=\{\bot,1,2,3\}$ and $\D=\<\I,\IC,{P}\>$ be a deductive
database, where $\I=\<\{p(\bot)\},\emptyset\>$, $\IC=\{p(2)\rightarrow
\bott\}$ and ${P}=\{q(X) \leftarrow p(X)\}$.

We have $\W(\I,\IC)=\{\{p(1)\},\{p(3)\},$ $\{p(1),p(3)\}\}$. Each of
the possible worlds in $\W(\I,\IC)$, when extended with the view
${P}$, gives rise to a program that has answer sets. Thus,
$\D$ is consistent. Moreover, the possible worlds for $\D$ are
$\{p(1),q(1)\}$, $\{p(3),q(3)\}$ and $\{p(1),q(1),p(3),q(3)\}$ (in
this case, one for each possible world in $\W(\I,\IC)$).
It follows that $p(\bot)$ and $q(\bot)$ are true, $p(2)$ and $q(2)$
are false, and $p(1),q(1),p(3)$ and $q(3)$ are unknown in $\D$.
%Consequently, the sets of \emph{true}, \emph{false} and
%\emph{unknown} facts are: $\D_t=\{p(\bot),q(\bot)\}$,
%$\D_f=\{p(2),q(2)\}$ and $\D_u=\{p(1),q(1),p(3),q(3)\}$.
\hfill$\Box$
\end{example}

\section{View Updating}
\label{vus}

In the \emph{view update problem}, the user specifies a \emph{request},
a list of facts the user learned (observed) to be true or false,
and wants the stored database to be updated to reflect it.\footnote{We do
not allow requests that facts be \emph{unknown}. That is, we only allow
definite requests. While there may be situations when all the user
learns about the fact is that it is unknown, they seem to be rather rare.
In a typical situation, the user will learn the truth or falsity of a
fact.}

\begin{definition}%[\textsc{Request over a deductive database}]
\label{update-request}
A \emph{request} over a deductive database $\D$ is a pair
$\RS=(\RS_t,\RS_f)$, where $\RS_t$ and $\RS_f$ are
disjoint sets of facts requested to be true and false,
respectively.
\hfill$\Box$
\end{definition}

To fulfill a request we need an update which, when executed,
yields a database such that the view it determines satisfies the request.
We call such updates \emph{weak repairs}.

\begin{definition}%[\textsc{Weak Repair}]
Let $\D=\<\I,\IC,{P}\>$ be a deductive database and $\RS$ a request.
An update ${U}$ for $\I$ is a \emph{weak repair} for ($\D,\RS)$
if ${U}$ \emph{fulfills} $\RS$, that is, if for every $a\in\RS_t$,
$v_{\cal{D}\circ{U}}(a)=\tr$ and for every
$a\in\RS_f$, $v_{\cal{D}\circ{U}}(a)=\fa$. \hfill$\Box$
\end{definition}

We are primarily interested in updates that do not drastically change
the database. One condition of being ``non-drastic'' is not to introduce new
predicate or constant symbols. That leads us to the notion of a relevant
weak repair.

\begin{definition}%[\textsc{Relevant Weak Repair}]
Let $\D=\<\I,\IC,{P}\>$ be a deductive database and $\RS$ a request.
A constant is
\emph{relevant with respect to $\D$ and $\RS$} if it occurs in $\D$,
or $\RS$, or if it is $\bot$. A predicate is \emph{relevant with respect
to $\D$ and $\RS$} if it occurs in $\D$ or in $\RS$.
 A weak repair ${U}$ for $(\D,\RS)$ is \emph{relevant} if every
constant and predicate occurring in ${U}$ is \emph{relevant}.
\hfill$\Box$
\end{definition}

More generally, a weak repair is ``non-drastic'' if it minimizes the change
it incurs \cite{Todd77}. There are two aspects to the minimality of change:
(1) minimizing the set of new predicate symbols and constants introduced by
an update to the database (in the extreme case, no new symbols must be
introduced, and we used that requirement to define relevant weak repairs
above); (2) minimizing the change in the truth values of facts with respect
to the database. Following the \emph{Ockham's Razor} principle to avoid
introducing new entities unless necessary, we take the minimality of the
set of new symbols as a primary consideration. To define the resulting
notion of change minimality, we assume that the truth values are ordered
$\fa\leq\un\leq\tr$. Further, for a deductive database $\D$, a
request set $\RS$ and an update ${U}\in \U$ we define $NC(\D,\RS,{U})$ as
the set of non nullary constants that occur in ${U}$ and not in $\D$ and $\RS$.
%that is, the set of \emph{new constants} introduced by ${U}$ with respect to
%the ones appearing in $\D$ and $\RS$.

\begin{definition}
\label{minimality}
Let $\D=\<\I,\eta\>$ be a database with integrity constraints. For updates
${V},{U}\in\U$, we define ${U} \sqsubseteq{V}$ if:
$NC(\D,\RS,{U})\subset NC(\D,\RS,{V})$, or
$NC(\D,\RS,{U})= NC(\D,\RS,{V})$ and for every \emph{base atom} $a$
\begin{enumerate}
\item if $v_{\cal{D}}(a)=\tr$, then $v_{\cal{D}\circ{U}}(a) \geq v_{\cal{D}\circ{V}}(a)$
\item if $v_{\cal{D}}(a)=\fa$, then $v_{\cal{D}\circ{V}}(a) \geq v_{\cal{D}\circ{U}}(a)$
\item if $v_{\cal{D}}(a)=\un$, then $v_{\cal{D}\circ{U}}(a)=\un$ or
$v_{\cal{I}\circ{V}}(a) = v_{\cal{D}\circ{U}}(a)$.
\end{enumerate}
We also define ${U}\sqsubset{V}$ if ${U}\sqsubseteq{V}$ and ${V}\not\sqsubseteq{U}$.\hfill$\Box$
\end{definition}

We now define the classes of \emph{repairs} and \emph{relevant repairs} as
subclasses of the respective classes of weak repairs consisting of their
$\sqsubseteq$-minimal elements.

\begin{definition}%[\textsc{Repairs and Relevant Repairs}]
Let $\D=\<\I,\IC,{P}\>$ be a deductive database and $\RS$ a request.
A (\emph{relevant}) \emph{repair} for $(\D,\RS)$ is a $\sqsubseteq$-minimal
(relevant) weak repair for $(\D,\RS)$.
\hfill$\Box$
\end{definition}

We note that the existence of (weak) repairs does not guarantee the
%\mc{Here and in other places, notation for updates has to be made consistent
%with definitions}
existence of relevant (weak) repairs.
The observation remains true even if the view is empty.

\begin{example}Let
$\D=\<\I,\IC,{P}\>$, where $\I=\<\emptyset,\emptyset\>$, $\eta=\emptyset$
and ${P}=\{t\leftarrow p(x),q(x)\}$. If the request is $(\{t\},\emptyset)$,
then each repair is of the form $\{+p(i)^D,$ $+q(i)^D\}$, for some $i
\in\Dom_d$. None of them is relevant. (We note that $\{+p(\bot)^D,+q(\bot)^D\}$
is not a (weak) repair. The database resulting from the update would admit
possible worlds of the form $\{p(i),q(j)\}$, where $i\not=j$. Clearly, the
corresponding possible world of the view over any such database does not
contain $t$ and so the update does not fulfill the request.)\hfill$\Box$
%If in this example $\I=\{s(1)\}$ then
%the set of repairs remains the same. One of these
%repairs, $\{+p(1)^D,+q(1)^D\}$, is relevant.
\end{example}

Some relevant constants are not ``forced'' by the database and the request,
that is, can be replaced by other constants. If such constants are
present in a relevant (weak) repair, this repair is \emph{arbitrary}.
Otherwise, it is \emph{constrained}. A formal definition follows.

\begin{definition}%[\textsc{Constrained (Weak) Repairs}]
\label{def-constrained-alt}
Let $\D=\<\I,\IC,{P}\>$ be a deductive database and $\RS$ a request.
A relevant (weak) repair ${U}$ for $(\D,\RS)$ is \emph{constrained} if
there is no non-nullary constant $a$ in ${U}$ such that replacing some
occurrences of $a$ in ${U}$ with a constant $b \neq a$ ($b$ might be $\bot$),
results in a weak repair for $(\D,\RS)$.
\hfill$\Box$
\end{definition}

\begin{example}\label{ex-occurrences}
Let %$\Dom=\{\bot,1,2,\ldots\}$ and
$\D=\<\I,\IC,{P}\>$, where $\I=
\<\{p(1),h(2)\}, \emptyset\>$, $\IC=\emptyset$ and ${P}=\{t\leftarrow
p(X), q(X); \ \ s\leftarrow r(X)\}$. Let us consider the request $\RS=
(\{s,t\},\emptyset)$. The updates $\R_i=\{+q(1)^D,+r(i)^D\}$, $i\in\{\bot,1,2\}$, and
$\R'_{i}=\{+q(2)^D,+p(2)^D,+r(i)^D\}$, $i\in\{\bot,1,2\}$, are relevant weak
repairs. One of them, $\R_\bot$, is constrained. Indeed, replacing in
$\R_\bot$ the unique occurrence of a non-nullary constant (in this
case, 1) with any other constant does not yield a weak repair. On the other
hand,
$\R_i$, $i\in\{1,2\}$, and $\R'_i$, $i\in\{\bot,1,2\}$, are not constrained.
Indeed, replacing with 3 the second occurrence of $1$ in $\R_1$, or the
occurrence of $2$ in $R_2$, or both occurrences of $2$ in $\R_i'$ in each
case results in a weak repair. Also weak repairs $\R_i=\{+q(1)^D,+r(i)^D\}$ and
$\R'_{i}=\{+q(2)^D,+p(2)^D,+r(i)^D\}$, $i\in\{3,\ldots\}$, are not constrained
as they are not even relevant.
%Indeed,
%replacing $i$ with itself ($i$ is non-relevant) results in a weak repair.
\hfill$\Box$
\end{example}

We stress that, in order to test whether a relevant (weak) repair $\R$ is
constrained, we need to consider every \emph{subset of occurrences} of
non-nullary constants in $\R$. For instance, in the case of the repair
$\R_1=\{q(1),r(1)\}$ from Example \ref{ex-occurrences}, the occurrence
of the constant $1$ in $q(1)$ is constrained by the presence of $p(1)$.
Replacing that occurrence of 1 with $3$ does not result in a weak
repair. However, replacing the occurrence of $1$ in $r$ with $3$ gives
a weak repair and shows that $\R_1$ is not constrained.

\begin{example}\label{ex-occurrences2}
Let $\Dom=\{\bot,1,2,\ldots\}$ and $\D=\<\I,\IC,{P}\>$, where $\I=
\<\{q(1,2),s(1,2,3)\},$ $ \emptyset\>$, $\IC=\emptyset$ and ${P}=\{p(X) \leftarrow
q(X,Y), r(X,Y,Z); \ \ r(X,Y,Z) \leftarrow s(X,Y,Z),t(X,Y,Z)\}$. Let us consider the request $\RS=
(\{p(1)\},\emptyset)$. In this case, our approach yields a unique constrained repair
$\R = \{+t(1,2,3)^D \}$. It recognizes that thanks to $s(1,2,3)$ simply
inserting $t(1,2,3)$ guarantees $r(1,2,3)$ to be true and, consequently, ensures the
presence of $p(1)$ in the view. There are other repairs and other relevant
repairs, but only the one listed above is constrained.
\hfill$\Box$
\end{example}

We observe that every (relevant, constrained)
weak repair contains a (relevant, constrained) repair.

\begin{proposition}
\label{prop:existence}
Let $\D=\<\I,\IC,{P}\>$ be a deductive database and $\RS$ a request.
A (relevant, constrained) repair for $(\D,\RS)$ exists if and only if
a (relevant, constrained) weak repair for $(\D,\RS)$ exists.\hfill$\Box$
\end{proposition}

\vspace*{-0.2in}
\section{Complexity}
\label{cs}

Finally, we discuss the complexity of decision problems concerning
the existence of (weak) repairs of types introduced above. The results we
present here have proofs that are non-trivial despite rather strong assumptions
we adopted. We present them in the appendix.

We assume that
the sets of base and derived predicate symbols, the set of integrity
constraints $\eta$ and the view ${P}$ are fixed. The only varying parts
in the problems are a database $\I$ and a request $\RS$. That is,
we consider the \emph{data complexity} setting. Moreover, we assume
that $\Dom=\{\bot,1,2,\ldots\}$, and take $=$ and $\leq$, both with the
standard interpretation on $\{1,2,\ldots\}$, as the \emph{only} built-in
relations.
We restrict integrity constraints to expressions of the form:
%\[
%\exists X( \forall Y (A_1\land\ldots\land A_k \rightarrow B_1\lor\ldots\lor
%B_m))\ \ \ (1)
%\]
$\exists X( \forall Y (A_1\land\ldots\land A_k \rightarrow B_1\lor\ldots\lor B_m))$,
where $A_i$ and $B_i$ are atoms with no occurrences of $\bot$ constructed
of base and built-in predicates, and where every variable occurring in
the constraint belongs to $X\cup Y$, and occurs in some atom $A_i$ built
of a base predicate.

We start by stating the result on the complexity of deciding the consistency
of an indefinite database with integrity constraints. While interesting in
its own right, it is also relevant to problems concerning the existence of
repairs, as one of the conditions for $U$ to be a repair is that the database
that results from executing $U$ be consistent.

\begin{theorem}
\label{prop:NP}
The problem to decide whether a database $\<\I,\eta\>$ has a possible world
(is consistent) is NP-complete.~\hfill$\Box$
\end{theorem}

We now turn attention to the problem of checking request satisfaction.
Determining the complexity of that task is a key stepping stone to the
results on the complexity of deciding whether updates are (weak) repairs
that are necessary for our results on the complexity of the
existence of (weak) repairs. However, checking request satisfaction turns
out to be a challenge even for very simple
classes of views. In this paper, we restrict attention to the case when ${P}$
is a safe definite (no constraints) acyclic (no recursion) Horn program,
although we obtained Proposition \ref{prop:horn1} in a more general form.
%We leave extending the results we present below to arbitrary Horn and
%stratified views as an open problem for the future work.

\begin{proposition}
\label{prop:horn1}
The problem to decide whether a ground atom $t$ is true in a deductive database
$\<\I,\eta,{P}\>$, where $\<\I,\eta\>$ is consistent and ${P}$ is a safe Horn
program, is in the class co-NP.~\hfill$\Box$
\end{proposition}

Next, we consider the problem to decide whether a ground atom $t$ is
false in a deductive database $\<\I,\eta,{P}\>$. We state it separately
from the previous one as our present proof of that result requires
the assumption of acyclicity.

\begin{proposition}
\label{prop:horn2}
The problem to decide whether a ground atom $t$ is false (ground literal
$\neg t$ is true) in a deductive database $\<\I,\eta,{P}\>$, where $\<\I,
\eta\>$ is consistent and ${P}$ is an acyclic Horn program, is in the class
co-NP.~\hfill$\Box$
\end{proposition}

With Propositions \ref{prop:horn1} and \ref{prop:horn2} in hand, we move
on to study the complexity of the problems of the existence of weak repairs.
First, we establish an upper bound on the complexity of checking whether
and update is a (relevant) weak repair.

\begin{proposition}
\label{checkWR-proposition}
Let $\D=\<\I,\IC,{P}\>$, where $\IC$ is a set of integrity constraints,
${P}$ an acyclic Horn program, ${U}$ an update and $\RS$ a request set.
The problem of checking whether an update ${U}$ is a weak repair (relevant
weak repair) for $(\D,\RS)$ is in $\Delta^P_2$.~\hfill$\Box$
\end{proposition}

With the results above, we can address the question of the complexity of
the existence of repairs. The first problem concerns weak repairs and stands
apart from others. It turns out, that deciding the existence of a weak repair
is NP-complete, which may seem at odds with Proposition
\ref{checkWR-proposition} (an obvious non-deterministic algorithm guesses
an update $U$ and checks that it is a weak repair apparently performing a
``$\Sigma_2^P$ computation''). However, this low complexity of the problem
is simply due to the fact that there are no relevance, constrainedness or
minimality constraints are imposed on weak repairs.
Thus, the question can be reduced to the question whether there
is a ``small'' database $\J$, in which the request holds. The corresponding
weak repair consists of deleting all elements from $\I$ and ``repopulating''
the resulting empty database so that to obtain $\J$.

\begin{theorem}
\label{existsWR}
Let $\D=\<\I,\IC,{P}\>$, where $\IC$ is a set of integrity constraints,
and ${P}$ an acyclic Horn program, and let $\RS$ be a request set.
The problem of deciding whether there is a weak repair
for $(\D,\RS)$ is $NP$-complete.~\hfill$\Box$
\end{theorem}

As noted, the case of the existence of weak repairs is an outlier and
deciding the existence of (weak) repairs of other types is much harder
(under common assumptions concerning the polynomial hierarchy).

\begin{theorem}
\label{existsRR}
Let $\D=\<\I,\IC,{P}\>$, where $\IC$ is a set of integrity constraints,
and ${P}$ an acyclic Horn program, and let $\RS$ be a request set.
The problems of deciding whether there is a relevant weak repair
and whether there is a relevant repair
for $(\D,\RS)$ are $\Sigma^P_2$-complete.~\hfill$\Box$
\end{theorem}

The last result concerns constrained (weak) repairs. It provides an upper
bound on the complexity of the problem of deciding the existence of
constrained repairs. We conjecture that the upper bound is in fact
tight but have not been able to prove it. We leave the problem for future
work.

\begin{theorem}
\label{existsCR}
Let $\D=\<\I,\IC,{P}\>$, where $\IC$ is a set of integrity constraints,
and ${P}$ an acyclic Horn program, and let $\RS$ be a request set.
The problems of deciding whether there is a constrained weak repair and
whether there is a constrained repair
for $(\D,\RS)$ are in $\Sigma^P_3$.~\hfill$\Box$
\end{theorem}

\section{Discussion and conclusion}

We presented a declarative framework for view updating and integrity
constraint maintenance for indefinite databases. The framework is based on
the notion of an indefinite deductive database. In our approach, the
indefiniteness appears in the extensional database and is modeled by a
single null value, consistent with the standards of database practice (a
condition not followed by earlier works on the view-update problem over
indefinite databases).  We defined a precise semantics for indefinite
deductive databases
in terms of possible worlds. We used the framework to formulate and study
the view-update problem. Exploiting the concept of minimality of change
introduced by an update, we defined several classes of repairs, including
relevant and constrained repairs, that translate an update request against
a view into an update of the underlying database. Finally, we obtained several
complexity results concerning the existence of repairs.

Our paper advances the theory of view updating in three main ways. First,
it proposes and studies the setting where extensional databases are
indefinite both before and after an update. While \emph{introducing}
indefiniteness to narrow down the class of potential repairs was
considered before \cite{Ten*2003}, the assumption there was that
the initial extensional database was complete. That assumption substantially
limits the applicability of the earlier results. Second, our paper proposes a more
expressive model of an indefinite extensional database. In our model
databases are determined by two sets of facts. The first set of facts
specifies what is true and provides an upper bound to what might still be
unknown. By CWA, everything else is false. The second
set of facts lists exceptions to the ``unknown range,'' that is, facts
that according to the first set might be unknown but are actually false
(exceptions). Third, our paper introduces two novel classes or repairs, relevant
and constrained, that often substantially narrow down possible ways to
fulfill an update request against a view. Relevant repairs do not introduce
any new constants and minimize change. Constrained repairs in addition do
not involve constants that are in some precise sense ``replaceable'' and,
thus, not grounded in the problem specification.

We already discussed some earlier work on view updating in the introduction
as a backdrop to our approach. Expanding on that discussion, we note that
the view-update problem is closely related to abduction and is often
considered from that perspective. Perhaps the first
explicit connection between the two was made by Bry \cite{Bry90} %\citeyear{Bry90}
who proposed
to use deductive tools as a means for implementing the type of abductive
reasoning required in updating views. That idea was pursued by others with
modifications that depended on the class and the semantics of the views.
For instance, Kakas and Mancarella \cite{kakas90} exploited in their %\citeyear{kakas90} exploited in their
work on view updates the abductive framework by Eshghi and Kowalski
\cite{EshghiK89}
%\citeyear{EshghiK89}
and were the first to consider the stable-model
semantics for
views. Neither of the two works mentioned above was, however, concerned
with the case of updates to views over indefinite databases. Console et
al. %\citeyear{cons95}
\cite{cons95} studied the case in which \emph{requests} can involve
variables.
These variables are replaced by null values and, in this way, null values
eventually end up in repaired databases. However, once there, they loose
their null value status and are treated just as any other constants.
Consequently, no reasoning over null values takes place, in particular,
they have no special effect on the notion of minimality. None of the
papers discussed studied the complexity of the view-update problem. Instead,
the focus was on tailoring resolution-based deductive reasoning tools to
handle abduction. Some results on the complexity of
abduction for logic programs were obtained by Eiter, Gottlob and Leone
%\citeyear{EiterGL97}. However, again the setting they considered did not assume
\cite{EiterGL97}. However, again the setting they considered did not assume
incompleteness in extensional databases.

Our paper leaves several interesting questions for future work. First, we
considered restricted classes of views. That suggests the problem
to extend our complexity results to the full case of Horn programs and,
later, stratified ones. Next, we considered a limited class of integrity
constraints. Importantly, we disallowed tuple-generating constraints.
However, once they are allowed, even a problem of repairing consistency in
an extensional database becomes undecidable. A common solution in the
database research is to impose syntactic restrictions on the constraints \cite{GrecoST11}.
%(such as weak acyclicity).
That suggests considering
view-updating in the setting in which only restricted classes of %so restricted tuple-generating
constraints are allowed.
%Finally, even under the restrictions we imposed,
%we only obtained an upper bound on the
%problem of the existence of constrained repairs. We conjecture
%that this upper bound is in fact tight.

%\bibliographystyle{acmtrans}

\section*{Acknowledgments}
%The first, second and fourth author were supported by XXXXX XXXXX XXXXX XXXXX
The third author was supported by the NSF grant IIS-0913459.
We are grateful to Sergio Greco and Leopoldo Bertossi for helpful discussions.

%\bibliographystyle{plain}
%\bibliography{biblio}

%\end{document}
\newpage
\section*{Appendix}

\smallskip
\noindent
\textbf{Proposition~\ref{prop:1}.}
\emph{Let $\I$ be a
database and $q$ a fact. Then $q\in\I_t$ if and only if
$W\models q$, for every $W\in\W(\I)$, and $q\in\I_f$ if and
only if $W\not\models q$, for every $W\in\W(\I)$.
}

\smallskip
\noindent Proof:
(1) Let $q\in\I_t$ and $W\in\W(\I)$. By Definition
\ref{poss-world}, we have that $q\in W^\Downarrow$. Consequently,
$W\models q$. Conversely, let us suppose that for some atom
$q$, $W \models q$, for every $W \in \W(\I)$. We will show that
$q\in\I^\Downarrow$. To this end, let us assume that
$q\notin\I^\Downarrow$.

Let $q'\in\I$. We will show that there is a definite fact $q_d$
such that $q' \preceq q_d$ and $q \not\preceq q_d$. It is
evident if $q'$ and $q$ are based on a different predicate
symbol. Thus, let as assume that $q'= p(s_1,\ldots,s_k)$ and
$q=p(t_1,\ldots,t_k)$, where all $s_i$ and $t_i$ are constants
from $\Dom$. If there is $i$, $1\leq i\leq k$, such that
$s_i\not=\bot$, $t_i\not=\bot$ and $s_i\not=t_i$, the claim is
again evident. Thus, let us assume that for every $i$, $1\leq
i\leq k$, $s_i= \bot$, or $t_i=\bot$ or $s_i=t_i$. As
$q\not\preceq q'$, there is $i$, $1\leq i\leq k$, such that
$s_i=\bot$ and $t_i=c$ for some $c\in\Dom \setminus\{\bot\}$.
We define $q_d$ to be any definite fact of the form
$p(u_1,\ldots,u_k)$ such that $q'\preceq q_d$ and $u_i\not=c$
(such facts exist as $\Dom$ is infinite). Clearly,
$q\not\preceq q_d$ and the claim follows.

We define $V=\{q_d\st q'\in\I\}$. By the definition,
$\I\subseteq V^\Downarrow$. Thus, $\I_t=\I^\Downarrow\subseteq
V^\Downarrow$. Furthermore, by the way atom $q_d$ were selected,
$V\subseteq\I^\Uparrow$. It follows that $V$ is a possible
world for $\I$. On the other hand, $V\not\models q$, a
contradiction.

\smallskip
\noindent (2) Let $q\in\I_f$ and let $W\in \W(\I)$. If
$W\models q$, then there is a definite atom $q_d\in W$ such
that $q\preceq q_d$. Since $W\subseteq\I^\Uparrow$, there is
an element $q'\in\I$ such that $q'\preceq q_d$. It follows that
$q\approx q'$ and so, $q\in\I^\approx$, a contradiction. Thus,
$W\not\models q$.

Conversely, let us consider an atom $q$ such that $W\not\models
q$, for every $W\in\W(\I)$. Let us suppose that
$q\in\I^\approx$. Then, there is a definite fact $q_d \in
\I^\Uparrow$ such that $q \preceq q_d$. Let $V$ be the set of
all definite facts in $\I^\Uparrow$. Clearly, $\I_t\subseteq
V^\Downarrow$ and $V\subseteq\I^\Uparrow$. Thus, $V$ is a
possible world for $\I$. Moreover, $q_d\in V$. Thus, $V\models
q$, a contradiction~\hfill$\Box$
\\

%\noindent
%\emph{Proposition~\ref{prop:2}}\\
%Let $\I$ and $\I'$ be databases. There is an update ${U}\in\U$ such that
%$\I\circ {U}=\I'$.~\hfill$\Box$
%\\

\noindent
\textbf{Proposition~\ref{prop:existence}.}
\emph{Let $\D=\<\I,\IC,{P}\>$ be a deductive database and $\RS$ a request.
A (relevant, constrained) repair for $(\D,\RS)$ exists if and only if
a (relevant, constrained) weak repair for $(\D,\RS)$ exists.}

\smallskip
\noindent Proof: The ``only if'' part is evident. By the definition, a repair is
a weak repair and, similarly, a relevant (constrained, respectively) repair
is a relevant (constrained, respectively) weak repair.

For the ``if'' part, let us consider a weak repair ${U}$ for $(\D,\RS)$.
If ${V}$ is a weak repair for $(\D,\RS)$ and ${V}\sqsubseteq {U}$, then
$NC(\D,\RS,{V})\subseteq NC(\D,\RS,{U})$. It follows that the family of all
weak repairs ${V}$ such that ${V}\sqsubseteq {U}$ is finite. Thus, it has a
minimal element, that is, it contains a repair for $(\D,\RS)$. The argument
for relevant (constrained, respectively) weak repairs is similar.~\hfill$\Box$

\paragraph{Complexity results. }
In order to prove Theorem~\ref{prop:NP} we need some auxiliary definitions and results.
We write $size(S)$, for the size of the representation
of $S$, where $S$ stands for a deductive database, database, set of integrity
constraints, etc.

%\noindent
\begin{lemma}\label{lem:cons}
If $\<\I,\eta\>$ is consistent, then it has a possible world $U$ such
that $|Dom(U)|=O(size(\I)+size(\eta))$.
\end{lemma}

\smallskip
\noindent Proof: If $\eta$ is empty then the claim trivially follows.
Let us assume that $\eta\neq \emptyset$. Let $W$ be a possible world for $\<\I,\eta\>$. We denote by $C$
the set of elements from $\Dom_\df$ that occur in $\<\I,\eta\>$.
Let us consider a constraint $\Phi\in\eta$. It is of the form
\[
\Phi = \exists X \forall Y \vph,
\]
where $\vph$ contains no quantifiers (we recall that $\vph$ is subject
to the syntactic restrictions we imposed on integrity constraints),
and $X\not=\emptyset$. Let us assume that $X=\{X_1,\ldots, X_j\}$. Since
$\Phi$ holds in $W$, we can select constants $c^\Phi_1,\ldots, c^\Phi_j\in
Dom(W)$ so that the constraint
\[
\Phi' = \forall Y \vph(X_1/c_1,\ldots, X_t/c_j)
\]
holds in $W$. Let $C'$ be the set of all constants of the form $c^\Phi_i$,
where $\Phi$ ranges over all constraints in $\eta$ that contain existential
quantifiers. Clearly, $|C'|=O(size(\eta))$.

Let $\eta'$ be obtained from $\eta$ by replacing each constraint $\Phi=\exists
X \forall Y \vph$, where $\vph$ is quantifier-free and $X\not=\emptyset$, with
the constraint $\Phi'$. It is clear that $\eta'$ consists of universally
quantified sentences and that $W$ is a possible world for $\<\I,\eta'\>$
(indeed, $W$ is a possible world for $\I$ and $\eta'$ holds in $W$).

Let us now select a possible world $V$ for $\<\I,\eta'\>$ so that
$|Dom(V)\setminus (C\cup C')|$ is minimal. As $W$ is a possible world
for $\<\I,\eta'\>$, $V$ is well defined.

Let $c\in Dom(V)\setminus (C\cup C')$ and let $V'$ be the set of all
atoms in $V$ except for those that contain an occurrence of $c$. We
have that $\eta'$ holds in $V'$. Indeed, let $\Psi\in \eta'$, say
$\Psi = \forall Y \vph$, where every variable in $\vph$ belongs to $Y$
(and $\vph$ is an implication of the form specified above). Let $\vph'$
be a ground instance of $\vph$ (over $Dom_\df$). If $c$ does not occur
in the consequent of $\vph'$, then the truth values of the consequent
of $\vph'$ in $V$ and $V'$, respectively, are the same, while the truth
value of the antecedent can only change from true to false when we move
from $V$ to $V'$. Consequently, the truth value of $\vph'$ remains true
when we move from $V$ to $V'$. If $c$ occurs in the consequent of $\vph'$,
then it also occurs in the base atom in the antecedent of $\vph'$. Thus,
the truth value of the antecedent of $\vph'$ is false in $V'$ ($V'$
contains no atoms with an occurrence of $c$) and so, $\vph'$ is true in
$V'$.

It follows that $V'$ is not a possible world for $\I$ (if it were, it
would be a possible world for $\<\I,\eta'\>$ with $|Dom(V')\setminus
(C\cup C') |< |Dom(V)\setminus (C\cup C')|$, a contradiction with the
way $V$ was chosen). Let $\I=\<D,E\>$. Since $V'$ was obtained from $V$
by deleting some of its atoms, every atom in $E$ is false in $V'$ as it
was false in $V$. It follows that there is an atom $A\in D$ such that
$A$ is not true in $V'$. Since $V$ is a possible world of $\I$, $V\models A$.
Let us assume that $A$ is definite. Then $A\in V$ and all constants it
contains belong to $C$. Thus, $A\in V'$ and $V'\models A$, a contradiction.
It follows that $A$ contains occurrences of $\bot$. We assign to $c$ that
atom $A$. Let $k$ be the number of occurrences of $\bot$ in $A$. Then, at
most $k$ constants in $Dom(V)\setminus (C\cup C')$ can be assigned $A$.
Indeed, in order for $A$ to be assigned to a constant $c\in Dom(V)\setminus
(C\cup C')$, every definite instantiation of $A$ in
$V$ must contain an occurrence of $c$ (otherwise, removing all atoms
with an occurrence of $c$ from $V$ would yield a possible world $V$ in
which $A$ would be true). Let $A'$ be any instantiation of $A$ in $V$.
If any of the constants assigned $A$ does not occur in $A'$, removing that
constant does not result in a possible world in which $A$ is false. Thus,
every constant assigned in $A$ must occur in $A'$, and the claim follows.
Let $a$ be the largest arity of a base predicate. Then, the number of
constants in $|Dom(V)\setminus (C\cup C')|$ is at most $|D|\times a=O(size(\I))$
(as $a$ is fixed, due to the fact that the database schema is fixed). It
follows that $|Dom(V)| \leq |Dom(V)\setminus (C\cup C')| + |C| + |C'| =
O(size(\I)+size(\eta))$. \hfill$\Box$
\\

\noindent
\begin{proposition}\label{prop:cons}
If $\<\I,\eta\>$ is consistent, then it has a possible world $V$ such
that $size(V)=O(pol($ $size(\I)+size(\eta)))$, where $pol$ is some polynomial
independent of $\<\I,\eta\>$.
\end{proposition}
Proof: Let $V$ be a possible world for $\<\I,\eta\>$ constructed in the
proof of Lemma \ref{lem:cons}. By Lemma \ref{lem:cons}, $|V| \leq b\times
((size(\I)+size(\eta))^a$, where $b$ is the number of base predicate
symbols and $a$ is the maximum arity of a base predicate (both numbers
are fixed and independent of $\<\I,\eta\>$).
Without loss of generality, we can assume that all integers in $V$ that
do not appear in $\<\I,\eta\>$ are not larger than $Max+O(size(\I)+
size(\eta))$, where $Max$ is the largest element from $\Dom_d$ occurring
in $\I$ or $\eta$, or 0, if $\<\I,\eta\>$ contains no constants. Indeed,
because of our restriction on built-in predicates, the only thing that
matters is how those integers relate to each other with respect to $=$
and $\leq$. Thus, all ``gaps'' among those integers in $\D(V)$ that
are larger than $Max$ can be eliminated by ``shifting'' the numbers to
the left. It follows that we can assume that the largest integer occurring
in $V$ is of size $O(size(\I)+size(\eta))$ and so, $size(V)\leq |V|\times
a \times O(size(\I)+size(\eta)) = O((size(\I)+size(\eta))^{a+1})$.
Since $V'$ is a possible world for $\<\I,\eta'\>$ it is also a possible
world for $\<\I,\eta\>$. Thus, the assertion follows. \hfill$\Box$

\smallskip
For the argument above it is important that the database schema is fixed.
For instance, if $\I=\<\{p(\bot,\ldots,\bot),q(\bot),r(\bot)\},\emptyset\>$,
where the arity of $p$ is $n$ and is not fixed, and $W= \{q(1), r(2)\}\cup
\{p(a_1,\ldots,a_n)\st a_i\in\{1,2\}\}$, then no constant can be removed
from $W$, $size(\I)=O(n)$ and $size(W)=O(n2^n)$.
\\

\noindent
\textbf{Theorem~\ref{prop:NP}.}
\emph{The problem to decide whether a database $\<\I,\eta\>$ has a possible world
(is consistent) is NP-complete.
}

%\begin{proof}
\smallskip
\noindent Proof:
(Membership) Proposition \ref{prop:cons} asserts that if $\<\I,\eta\>$ is
consistent, then it has a possible world whose size is polynomial in
$size(\I)+size(\eta)$. Thus, the following non-deterministic procedure
decides the problem: First, the procedure guesses a set of facts $W$ of size
implied by the constructions from Lemma \ref{lem:cons} and Proposition
\ref{prop:cons} (which is polynomial in $size(\I)+size(\eta)$). Next, the
procedure checks that $W$ is a possible world for $\I$ and that it satisfies
all constraints in $\grnd(\eta)$. Since both these tasks can be accomplished
in polynomial time, the problem is in the class NP.

\smallskip
\noindent
(Hardness) The NP-hardness follows by a reduction from the 3-SAT problem
to decide whether a quantified boolean formula $\Phi=\exists X\; \varphi$
is \emph{true}, where $X$ is a set of propositional variables and $\varphi$
is a conjunction of 3-literal clauses over the set of atoms in~$X$.

Given such a formula $\Phi$, we assume that the domain $\Dom$ contains symbols
$true$ and $false$ for representing the truth values, as well as symbols for
representing atoms and clauses of the formula $\vph$. With some abuse of the
notation, we denote the sets of those symbols with $X$ and $Cl$, respectively.
We assume that all those symbols are distinct and different from the domain
elements $1$, $2$ and $3$, which we use to represent positions of literals
in clauses (for each clause we fix the order).

Further, we assume that we have base predicate symbols %$at$, $cl$, $pos$,
%$truthVal$,
$val$, $sat$, and $occur$. %The predicates $at$ and $cl$ are unary.
%They are used to identify atoms and clauses. The predicates $pos$ and
%$truthVal$ are also unary. They identify, respectively, the three positions,
%$1$, $2$ and $3$, of literals in clauses, and the two truth values $true$ and
%$false$.
The predicates $val$ and $sat$ are binary. We use $val$ to specify the truth
value of atoms in $X$ and $sat$ to specify the truth value of clauses in $Cl$
(given the assignment of truth values to atoms as determined by $val$).
Finally, the predicate $occur$ is 4-ary. If an atom $x$ occurs in a clause $c
\in Cl$ positively (negatively, respectively) in the position $p\in\{1,2,3
\}$, we represent that by an atom $occur(c,p,x,true)$ ($occur(c,p,x,false)$,
respectively). Clearly, all atoms of the form $occur(c,p,x,v)$ uniquely
determine the formula $\vph$. We define $\I=\<D,\emptyset\>$ where $D$ consists of:

\begin{enumerate}
%\item $at(x)$, for each $x\in X$
%\item $cl(c)$, for each $c\in Cl$
%\item $pos(p)$, for $p=1, 2, 3$
%\item $truthVal(true)$ and $truthVal(false)$
\item $val(x,\bot)$, for each $x\in X$
\item $sat(c,\bot)$, for each $c\in Cl$
\item $occur(c,p,x,true)$, if $x$ occurs positively in $c$ in a position
$p \in \{1,2,3\}$
\item $occur(c,p,x,false)$, if $x$ occurs negatively in $c$ in a position
$p \in \{1,2,3\}$.
\end{enumerate}

Next, we define $\eta$ to consist of the following constraints (we note all
of them satisfy the syntactic ``safety'' restriction we imposed):

\begin{enumerate}
%\item $\forall C,P,A,V\; (occur(C,P,A,V) \rra cl(C)$
%\item $\forall C,P,A,V\; (occur(C,P,A,V) \rra pos(P)$
%\item $\forall C,P,A,V\; (occur(C,P,A,V) \rra at(A)$
%\item $\forall C,P,A,V\; (occur(C,P,A,V) \rra truthVal(V)$
\item $\forall A,V\; (val(A,V)\rra V=true \lor V=false)$
\item $\forall A\; (val(A,true)\land val(A,false)\rra \bot)$
\item $\forall C,V\; (sat(C,V) \rra V=true \vee V=false)$
\item $\forall C\; (sat(C,true)\land sat(C,false)\rra \bot)$
\item $\forall C,P,A,V\; (sat(C,false)\land occur(C,P,A,V)\land val(A,V)\rra
\bot)$
\item $\forall C,A,A',A'',V,V',V''$\\
\phantom{xxxx}$(sat(C,true)\land\, occur(C,1,A,V)\land \neg val(A,V)$\\
\phantom{xxxx}$\land\, occur(C,2,A',V')\land \neg val(A',V')$\\
\phantom{xxxx}$\land\, occur(C,3,A'',V'')\land \neg val(A'',V'')\rra \bot)$
\item $\forall C\; (sat(C,false)\rra \bot)$.
\end{enumerate}

One can prove that $\Phi$ is true if and only if $\<\I,\eta\>$ is consistent.
\hfill$\Box$
\\

\noindent
\textbf{Proposition~\ref{prop:horn1}.}
\emph{The problem to decide whether a ground atom $t$ is true in a deductive database
$\<\I,\eta,{P}\>$, where $\<\I,\eta\>$ is consistent and ${P}$ is a safe Horn
program, is in the class co-NP.}

\smallskip
\noindent Proof: It suffices to show that the problem to decide whether $t$ is \emph{not}
true
in a deductive database $\<\I,\eta,{P}\>$ is in the class NP. That problem
consists of deciding whether there is a possible world $W$ of $\<\I,\eta\>$ such that $LM({P}\cup W)\not\models t$. If such a possible world $W$
exists, the construction from the proof of Proposition \ref{prop:cons} shows
that there is a possible world $V\subseteq W$ of the size polynomial in the
size of $\I$. Since $V\subseteq W$, we have $LM({P}\cup V)\not\models t$.
Thus, to decide that $t$ is not true in $\<\I,\eta,{P}\>$, one guesses $V$,
a polynomial-size possible world for $\<\I,\eta\>$, and then verifies that
it is a possible world for $\<\I,\eta\>$ and that $t$ does not hold in it.
These verification tasks are polynomial in the size of $\I$ and so, the
problem to decide whether $t$ is not true in a deductive database $\<\I,\eta,
{P}\>$ is in the class NP, as required.  \hfill$\Box$
\\

\noindent
\textbf{Proposition~\ref{prop:horn2}.}
\emph{The problem to decide whether a ground atom $t$ is false (ground literal
$\neg t$ is true) in a deductive database $\<\I,\eta,{P}\>$, where $\<\I,
\eta\>$ is consistent and ${P}$ is an acyclic Horn program, is in the class
co-NP.}

\smallskip
\noindent Proof: To prove the assertion, we show that the complementary problem to
decide whether $t$ is \emph{not} false in $\<\I,\eta,{P}\>$ is in NP. To
this end, it suffices to show that if $t$ is not false in $\<\I,\eta,{P}\>$
then there is a possible world $V$ of $\<\I,\eta\>$ such that the size of
$V$ is polynomial in the size of $\I$ and ${P}\cup V\models t$.

Clearly, if  $t$ is not false in $\<\I,\eta,{P}\>$, then there is a possible
world $W$ of $\<\I,\eta\>$ such that ${P}\cup W\models t$. Let $T$ be the tree
of a shortest resolution proof of $t$ from $ground({P}\cup W)$ and $W_T$ the
subset of atoms in $W$ occurring in $T$. Let $V$ be any subset of $W$ such
that
\begin{enumerate}
\item
$(W\cap\I_t)\cup W_T \subseteq V$
\item
$V$ is minimal with respect to that requirement.
\end{enumerate}
In other words, for $V$ we take any minimal subset of $W$ that still
``explains'' the truth of every true atom of $\I$ and contains all atoms $W_T$.
We stress that there are many of these sets. We have the following properties:
\begin{enumerate}
\item $LM({P}\cup V)\models t$ (holds, as $W_T\subseteq V$)
\item The size of $V$ is polynomial in the size of $\I$ (follows by the fact
that ${P}$ is not recursive and $T$ is a tree of a \emph{shortest} proof, which
implies that $|W_T|$ is bounded by a constant independent of the size of $\I$).
\end{enumerate}

As $V$ is a subset of $W$, it may happen that some integrity constraints
are violated. Specifically, it may now be that the body of an integrity
constraint is \emph{true} with respect to $W$, remains \emph{true} with
respect of $V$, and its head, which is \emph{true} in $W$, becomes is
\emph{false} in $V$. In order to ``fix'' that constraints we reinsert into
$V$ atoms from $W$ that make the consequent of the constraint true or,
to be precise, a minimal set of such atoms. In this way, we obtain a new set,
which we substitute for $V$. We continue this process as long as there are
violated integrity constraints. The key point is that this process can only
reintroduce a polynomial number of atoms because the integrity constraints
are safe (no existential quantification in the consequents) and the constants
involved in the process are only those occurring in $\I$ and in $T$.
Therefore, when the process terminates, $V$ is a possible world of $\<\I,\eta\>$, its size is polynomial in the size of $\I$ and
$LM({P}\cup V)\models t$. \hfill$\Box$
\\

We believe that the result above can be extended to the case of (non-acyclic)
Horn programs. However, a proof of such a result would have to follow a
different approach. The reason is that given a possible world $W$ for $\<\I,
\eta\>$ such that $LM({P}\cup W)\models t$, it is not always possible to
find a possible-world $V\subseteq W$ of size polynomial in the size of $\<\I,
\eta\>$ such that $LM({P}\cup V)\models t$. Let $\I=\<\{start(1), edge(\bot,
\bot)\},\emptyset\>$, $\eta=\emptyset$ and ${P}$ consist of:

\smallskip
\indent
$r(X) \leftarrow start(X)$\\
\indent
$r(X) \leftarrow r(Y), edge(Y,X), r(X)$.

\smallskip
\noindent
Let us define $t=r(0)$ and $W=\{start(1), edge(1,2), edge(2,3),
\ldots, edge(n-1,n),$ $edge(n,0)\}$. Then $W$ is a possible world for $\<\I,
\eta\>$ and $LM({P}\cup W)\models t$. But $W$ contains no fixed-size
possible world $V$ for $\<\I,\eta\>$ such that $LM({P}\cup V)\models t$.
Such a fixed-size possible world exists but has to be constructed
independently of any particular possible world.
\\

\noindent
\textbf{Proposition~\ref{checkWR-proposition}.}
\emph{Let $\D=\<\I,\IC,{P}\>$, where $\IC$ is a set of integrity constraints, ${P}$ an acyclic Horn program, ${U}$ an update and $\RS$ a request set.
The problem of checking whether an update ${U}$
is a weak repair for $(\D,\RS)$ is in $\Delta^P_2$.}

\smallskip
\noindent Proof:
Let $\D'=\<\I',\IC,{P}\>$, where $\I'=\I \circ{U}$. In order to test whether
${U}$ is a weak repair we need to check whether $\D'$ is consistent and
$\D'\models \RS$. Clearly, $\D'$ can be constructed in polynomial time. The
first problem is NP-complete and so, can be solved by a call to an NP oracle
(Theorem \ref{prop:NP}). The second one can be solved with a polynomial number
of calls to coNP oracles (cf. Propositions \ref{prop:horn1} and
\ref{prop:horn2}).  \hfill$\Box$
\\

\noindent
\textbf{Theorem~\ref{existsWR}.}
\emph{Let $\D=\<\I,\IC,{P}\>$, where $\IC$ is a set of integrity constraints,
and ${P}$ an acyclic Horn program, and let $\RS$ be a request set.
The problem of deciding whether there is a weak repair
for $(\D,\RS)$ is $NP$-complete.}

\smallskip
\noindent
Proof: \nop{
\textbf{MT: works only when no atom is requested to be unknown;
I am reasonably certain it can be generalized. The current membership proof also needs work.}}
(Membership)
Let us assume that there is a weak repair for $(\D,\RS)$. It follows that
there is an indefinite database $\J$ that satisfies the request. Let $W$
be a possible world of $\J$. Let $a$ be an atom required to be true. Since
$\J$ satisfies the request, there is a definite atom $b$ such that $a\preceq
b$ and $b$ has a proof from ${P}\cup W$. Without loss of generality we may
assume that the proof is minimal (no formula can be eliminated). It follows
that the cardinality of that proof does not depend on $\I$ nor on $\J$ nor
on $\RS$. It is so because ${P}$ is acyclic and fixed (it is independent of
the database component and of the request). Let $W_a$ be the set of atoms
from $W$ occurring in that proof, $C_a$ the set of all constants occurring
in that proof and let $W'$ and $C$ be the unions of all sets $W_a$ and $C_a$,
respectively, over all atoms $a$ required to be true. Finally, we define
$W''$ to be the set of all atoms in $W$ that contain only occurrences of
constants in $C$. It is clear that $W'\subseteq W''\subseteq W$ and that
the size of $W''$ is polynomial in the size of $\RS$. We will show that
the request $\RS$ holds in $W''$ and that $W''$ satisfies the integrity
constraints.

First, since $W'\subseteq W''$, ${P}\cup W''\models a$, for every atom $a$
requested to be true. Second, since for every atom $b$ requested to be false,
${P}\cup W\not\models b$, ${P}\cup W''\not\models b$. Both properties follow
from the fact that ${P}$ is a Horn program and $W'\subseteq W''\subseteq W$.

Next, we  note that $W''$ is obtained from $W$ by removing from $W$ all
atoms that contain at least one constant not in $C$. Since $W$ satisfies
all the integrity constraints and since for every ground instance of an
integrity constraint and for every constant that occurs in the consequent
there is a base predicate in the antecedent with an occurrence of that
constant, $W''$ satisfies all integrity constraints, as well.

To recap, we proved that if there is a weak repair for $(\D,\RS)$, then
there is a set of definite atoms $W''$ built of base predicates such that
the size of $W''$ is polynomial in the size of $\RS$, $W''$ satisfies
all the integrity constraints and ${P}\cup W''\models \RS$. Conversely,
if such a set $W''$ exists, then there is a weak repair for $(\D,\RS)$.
Indeed, such a repair might be constructed of all update atoms needed to
transform $\I$ into an empty database and all insert atoms converting
that empty database into $W''$. Since $W''$ is definite, $W''$ is the only
possible world of $\<W'',\eta,{P}\>$ and by the properties of $W''$, $\<W'',
\eta,{P}\>$ satisfies the request.

It follows that the following algorithm decides whether there is a weak
repair for $(\D,\RS)$: guess a polynomial-size definite database $W''$
and verify that it satisfies the integrity constraints and the request.
Both checks can be performed in polynomial time. Thus, the problem is in
the class NP.

\smallskip
\noindent
(Hardness) We prove the claim by a reduction from the 3-SAT problem. That
problem consists of deciding whether a formula $\Phi=\exists X\; \varphi$
is \emph{true}, where $X$ is a set of propositional atoms and $\varphi$ is
a CNF formula over $X$ with each clause being a disjunction of 3 literals.
We denote the set of clauses in $\vph$ by $Cl$.

Given such a formula $\Phi$, we assume that the domain $\Dom$ contains
symbols $true$
and $false$ to represent the truth values, and symbols to represent atoms and
clauses of $\vph$. With some abuse of notation we will denote those two sets
of symbols by $X$ and $Cl$, respectively. We also assume these symbols are
different from $1$, $2$ and $3$, which we will use to denote positions of
literals in clauses (we fix the order of literals in each clause). Next, we
assume that we have base predicates $val$ and $occur$. The predicate
$val$ is binary, and the predicate $occur$ is 4-ary. We use $val$ to define
the truth value of a propositional atom in $X$. If an atom $x$ occurs in a
clause $c$ positively (negatively,
respectively) in a position $p\in\{1,2,3\}$, we represent that by an atom
$occur(c,p,x,true)$ ($occur(c,p,x,false)$, respectively). As derived predicates,
we use predicates $sat$ and $occur'$. The predicate $sat$ is unary; we use it
to represent the truth value of clauses in $\varphi$. The predicate $occur'$
has the same arity and plays the same role as its non-primed counterpart. We
define $\I=\<\emptyset,\emptyset\>$ and $\eta$ to consist of the following constraints:

\begin{enumerate}
\item $\forall C,P,A,A'V\; (occur(C,P,A,V)\land occur(C,P,A',V)\rra A=A')$
\item $\forall C,P,A,V\; (occur(C,P,A,V)\rightarrow P=1 \vee P=2 \lor P=3)$
\item $\forall C,P,A,V\; (occur(C,P,A,V) \rra V=true \lor V=false$
\item $\forall C,P,A\; (occur(C,P,A,true)\land occur(C,P,A,false)\rra\bot)$
\item $\forall A,V\; (val(A,V)\rra V=true \lor V=false$
\item $\forall A\; (val(A,true)\land val(A,false) \rra \bot)$.
\end{enumerate}
Next, we define ${P}$ to consist of the following rules:

\begin{enumerate}
\item $occur'(C,P,W,V) \leftarrow occur(C,P,W,V)$
\item $sat(C) \leftarrow occur(C,P,W,V),val(W,V)$
\end{enumerate}
Finally, we define the request set $\RS$ to consist of the following literals:
\begin{enumerate}
\item $occur'(c,p,x,true) $, for every clause $c$ in $\vph$, $x\in X$ and
$p\in\{1,2,3\}$ such that $x$ occurs positively in $c$ in position $p$
\item $occur'(c,p,x,false) $, for every clause $c$ in $\vph$, $x\in X$ and
$p\in\{1,2,3\}$ such that $x$ occurs negatively in $c$ in position $p$
\item $sat(c)$, for every  $c \in Cl$.
\end{enumerate}
One can prove that $\Phi$ is \emph{true} if and only if a weak
repair for $(\D,\RS)$ exists. \hfill$\Box$

%\noindent
%\begin{corollary}~\label{checkWR-corollary}
%%\emph{Corollary~\ref{checkWR-corollary}}\\
%Let $\D=\<\I,\IC,{P}\>$, where $\IC$ is a set of integrity constraints,
%and ${P}$ an acyclic Horn program. Let also ${U}$ be an update and $\RS$ a
%request set.  The problem of checking whether an update ${U}$
%is a relevant weak repair for $(\D,\RS)$ is in $\Delta^P_2$.
%\end{corollary}
%Proof: Checking whether ${U}$ is a weak repair is in the class $\Delta_2^P$
%(Proposition \ref{checkWR-proposition}), and checking whether ${U}$ is relevant
%is a polynomial problem. Thus, the assertion follows. \hfill$\Box$

\medskip
\noindent
\textbf{Theorem~\ref{existsRR}.}
\emph{Let $\D=\<\I,\IC,{P}\>$, where $\IC$ is a set of integrity constraints,
and ${P}$ an acyclic Horn program, and let $\RS$ be a request set.
The problems of deciding whether there is a relevant weak repair
and whether there is a relevant repair
for $(\D,\RS)$ are $\Sigma^P_2$-complete.}

\smallskip
\noindent
Proof: \nop{MT: here we do not need to restrict to 2-valued requests}
Firstly observe that there are only finitely many relevant weak repairs. Therefore, a
$\sqsubseteq$-minimal relevant weak repair exists if and only if a relevant
weak repair exists. In the following we prove that the problem of deciding whether there is a relevant weak repair
for $(\D,\RS)$ is $\Sigma^P_2$-complete.

\smallskip
\noindent
(Membership) Each relevant weak repair for $(\D,\RS)$ consists of insert and
delete actions involving predicate symbols and constants occurring in $\D$ and
$\RS$ and the additional constant $\bot$ representing the null value. Therefore,
every relevant weak repair has size polynomial in the size of $(\D,\RS)$. Once
we guess a relevant update ${U}$, we have to test whether the updated database
$\D\circ {U}$ is consistent, which can be decided by an NP oracle (Theorem
\ref{prop:NP}), and whether the request is satisfied, which can be decided by
polynomially many calls to co-NP oracles (Propositions \ref{prop:horn1} and
\ref{prop:horn2}). Thus, the membership of the problem in the class $\Sigma_2^P$
follows.

\smallskip
\noindent
(Hardness) In order to show $\Sigma_2^P$-hardness, we present
a reduction from the problem to decide whether a quantified boolean formula
$\Phi=\exists X\forall Y\; \varphi$ is true, where $X$ and $Y$ are disjoint
sets of propositional atoms and $\varphi$ is a 3-DNF formula over the set of
atoms $X\cup Y$. It is indeed sufficient, as that problem is known to be
$\Sigma_2^P$-complete.

Given such a formula $\Phi$, we denote by $D_\vph$ the set of (identifiers
of) disjuncts in $\varphi$. Clearly, each $d\in D_\vph$ is a conjunction of
literals. We denote by $\varphi_d$ the set of atoms that occur in $d$.
We assume that the domain
$\Dom$ contains symbols $true$ and $false$, as well as symbols to represent
atoms in $X$ and in $Y$, and disjuncts of $\vph$ (elements of $D_\vph$).
With some abuse of notation we will denote those sets of symbols by $X$, $Y$
and $D_\vph$, respectively. We assume that all these symbols are different
from three additional domain symbols, $1$, $2$ and $3$. We denote the set of
all constants mentioned above by $\Dom_r$. That is, we set $\Dom_r=X\cup Y
\cup D_\vph\cup\{1,2,3\} \cup \{true,false\}$.

We assume base predicates $inX$, $inY$, $disj$, $occur$, $inX_c$, $inY_c$,
$disj_c$, $occur_c$, $val_X$, $val_Y$ and $assign$. The predicates $inX$,
$inY$ are unary and their extensions give the sets of atoms $X$ and $Y$.
The predicate $disj$ is also unary; its extension specifies the disjuncts of
the formula $\vph$. The predicate $occur$ is 4-ary. If an atom $a$ occurs in
a disjunct $d$ positively (negatively, respectively) in a position $p\in\{1,2,
3\}$, we represent that by an atom $occur(d,p,x,true)$ ($occur(d,p,x,false)$,
respectively). The predicates with the subscript $_c$ have the same arities
as their subscript-free counterparts and are intended to represent their
complements with respect to $\Dom_r$. The predicates $val_X$, $val_Y$ and
$assign$ are binary. We use $val_X$ and $val_Y$ to specify the truth value of
atoms in $X$ and $Y$,
respectively. The predicate $assign$ will associate some new constants (not
in $\Dom_r$) with $true$ and some other new constants with $false$ (the need
for that will became evident from the proof below).

We also assume derived predicates $q'$ for $q\in\{inX,inX_c,inY,inY_c,disj,
disj_c,occur,$ $occur_c\}$, each of the same arity as its non-primed
counterpart, as well as three additional ones
$val$, $sat$ and $satisfied$. The predicates $q'$, $q\in\{inX,inX_c,inY,inY_c,
disj,disj_c,occur,$ $occur_c\}$, allow us to formulate a request so that no
relevant weak repair can change the extensions of the corresponding predicates
in the database. The predicate $val$ is unary and allows us to formulate a
request that ensures that every possible world of the database resulting from a
relevant weak repair assigns a truth value to every atom in $X\cup Y$.
The $sat$ is unary and $satisfied$ is propositional (0-ary). We use them to
represent the satisfaction of individual disjuncts in $\vph$ and of $\vph$
itself.

We now define the database $\I=\<D,\emptyset\>$ where $D$ consists of the following atoms:

\begin{enumerate}
%\item $at(a)$, for every atom $a\in X\cup Y$
\item $inX(a)$, for every atom $a\in X$
\item $inY(a)$, for every atom $a\in Y$
\item $disj(d)$, for every disjunct $d\in D_\vph$
\item $occur(d,a,p,true)$ for every atom $a\in X\cup Y$ that occurs non-negated
in position $p$ in the conjunct $d$
\item $occur(d,a,p,false)$, for every atom $a\in X\cup Y$ that occurs negated
in position $p$ in the conjunct $d$
\item Finally, we include in $D$ atoms $inX_c(a)$, $inY_c(a)$, $disj_c(d)$,
$occur_c(d,a,p,v)$ so that the extensions of $inX_c$, $inY_c$, $disj_c$ and
$occur_c$ are the complements with respect to $\Dom_r$ ($\Dom_r^4$, in the
last case) of the extensions of the corresponding predicates as specified in
(1)-(5).
\end{enumerate}

We define $\eta$ to consist of the constraints:

\begin{enumerate}
\item $\forall A,V\; (val_X(A,V)\rra V=true \lor V=false)$
\item $\forall A\; (val_X(A,true)\land val_X(A,false)\rightarrow\bot)$
\item $\forall A,V\; (val_X(A,V)\rra inX(A))$
\item $\forall B,V\; (val_Y(B,V)\rra inY(B))$
\item $\forall B\; (val_Y(B,true)\rightarrow\bot)$
\item $\forall B\; (val_Y(B,false)\rightarrow\bot)$
\item $\forall B\; (val_Y(B,1)\rightarrow\bot)$
\item $\forall B\; (val_Y(B,2)\rightarrow\bot)$
\item $\forall B\; (val_Y(B,3)\rightarrow\bot)$
\item $\forall B,A\; (inX(A)\land val_Y(B,A)\rightarrow\bot)$
\item $\forall B,A\; (inY(A)\land val_Y(B,A)\rightarrow\bot)$
\item $\forall B,D\; (disj(D)\land val_Y(B,D)\rightarrow\bot)$
\item $\forall B,V,V'\; (val_Y(B,V)\land val_Y(B,V')\rra V=V')$
\item $\forall B,\; V(val_Y(B,V)\rra assign(V,true)\lor assign(V,false))$
\item $\forall B,V\; (val_Y(B,V)\land assign(V,true)\land assign(V,false)\rra\bot)$.
\end{enumerate}

The view ${P}$ consists of the following rules:
\begin{enumerate}
\item $q'(\textbf{W}) \lra q(\textbf{W})$, for every base predicate $q\in
\{inX,inX_c,inY,inY_c,disj,disj_c,occur,$ $occur_c\}$ (\textbf{W} stands for
a vector of different variable symbols of the appropriate arity)
\item $val(A) \lra inX(A),val_X(A,V)$
\item $val(A) \lra inY(A),val_Y(A,V)$
\item $sat(D,P) \lra occur(D,P,A,V),inX(A),val_X(A,V)$
\item $sat(D,P) \lra occur(D,P,A,V),inY(A),val_Y(A,V'),assign(V',V)$
\item $satisfied \lra sat(D,1),sat(D,2),sat(D,3)$.
%\item For each $d\in D_\vph$,\\
%$sat(c) \leftarrow \bigwedge_{x \in \varphi_c} (val_X(x,true)),
%\bigwedge_{\overline{x} \in \varphi_c} (val_X(x,false)),$\\
%\phantom{xxxxxxxx}$\bigwedge_{y\in \varphi_c} (val_Y(y,V_y), assign(V_y,true)),
%\bigwedge_{\overline{y}\in \varphi_c} (val_Y(y,V_y), assign(V_y,false))$
%\item
%$satisfied\leftarrow sat(C)$
\end{enumerate}

Finally, we define the request set $\RS_\Phi$ to consist of the facts:
\begin{enumerate}
\item $q'(\mathbf{a})$, for every atom of the form $q(\mathbf{a})$ in $\I$
($q\in \{inX,inX_c,inY,inY_c,disj,disc_c,$ $occur, occur_c\}$).
\item $val(A)$, for every $A\in X\cup Y$
\item $satisfied$.
\end{enumerate}

We note that the part (1) of the request guarantees that no matter what
relevant weak
repair we consider, the extensions of base predicates $inX$, $inX_c$,
$inY$, $inY_c$, $disj$, $disj_c$, $occur$ and $occur_c$ will be the same before
and after the repair is applied. Atoms given in the specification
(2) of the request guarantee that applying any weak repair results in a
database that contains $val_X(\bot,\bot)$ or, for every $a\in X$,
$val_X(a,true)$, $val_X(a,false)$ or $val_X(a,\bot)$. Since no constants
present in the original
database or in the request can be used as $\alpha$ in atoms of the form
$val_Y(y,\alpha)$, it also follows that every relevant weak repair inserts
into the database atom $val_Y(\bot,\bot)$ or, for every $b\in Y$, atoms
$val_Y(y,\bot)$. Due to the integrity constraint (14), it also inserts
$assign(\bot,\bot)$ or $assign(\bot,true)$ and $assign(\bot,false)$.

We now prove that there exists a relevant weak repair for the database
$(\D(\Phi),\RS(\Phi))$ if and only if
$\Phi=\exists X\forall Y\; \varphi$ is true. Let $W$ be a set of
propositional atoms. An \emph{assignment} $\Theta_W$ \emph{of truth values}
to atoms in $W$ determines a function $\Theta_W(\cdot)$, that takes as its
argument a formula $\varphi$ and returns the formula obtained by replacing
in $\varphi$ each $w\in W$ with a truth value (\emph{true} or \emph{false}).
Given two assignments  $\Theta_X(\cdot)$ and $\Theta_Y(\cdot)$, we define the
function $\Theta_{X,Y}(\cdot)=\Theta_X(\Theta_Y(\cdot))$.

\smallskip
\noindent
($\Rightarrow$)
Let ${U}$ be a relevant weak repair for the database $(\D(\Phi),\RS(\Phi))$. Our goal is to derive from ${U}$ an assignment $\Theta_X$ such
that for each assignment $\Theta_Y$ the truth value of $\Theta_{X,Y}(\vph)$ is
\emph{true}.

Let $\I'=\I\circ{U}$ and $x\in X$. By our discussion above, we have the following
cases:
\begin{enumerate}
\item  $\I'$ contains an atom of the form $val_X(x,true)$ ($val_X(x,false)$,
respectively). In this case we set $\Theta_X(x)=true$ ($\Theta_X(x)=false$,
respectively). We note that $\I'$ cannot contain both $val_X(x,true)$ and
$val_X(x,false)$ as that would violate one of the integrity constraints.
\item $\I'$ does not contain an atom of the form $val_X(x,true)$ or
$val_X(x,false)$. Then, $\I'$ contains $val_X(x,\bot)$ or $val_X(\bot,\bot)$.
In this case, the truth value of $X$ can be arbitrary and we set $\Theta_X(x)
=true$.
\end{enumerate}

We will now show that for each assignment $\Theta_Y$ of truth values to atoms
in $Y$, $\Theta_{X,Y}(\varphi)$ is \emph{true}. By our discussion above,
since ${U}$ is a relevant weak repair, $\I'$ contains the atom $val_Y(\bot,
\bot)$ or, for every $y\in Y$, the atom $val_Y(y,\bot)$, and it also contains
the atom $assign(\bot,\bot)$ or, the atoms $assign(\bot,true)$ and
$assign(\bot,false)$.

Let $\xi_t$ and $\xi_f$ be domain elements not in $\Dom_r$. We define
\[
\begin{array}{rcl}
W&=&\I\cup \{val_X(x,\Theta_X(x))\st x\in X\}\\
 &\cup& \{val_Y(y,\xi_t)\st y\in Y, \ \Theta_Y(y)=true\}\\
 &\cup& \{assign(\xi_t,true),assign(\xi_f,false)\}.
\end{array}
\]
It is straightforward to check that $W$ is a possible world of $\<\I',\eta\>$.
Thus, $satisfied$ holds in $LM({P}\cup W)$. Consequently, there is a disjunct
$d\in D_\vph$ such that $sat(d)$ holds in $LM({P}\cup W)$. It follows that
$\Theta_{X,Y}(\vph)=true$.

\smallskip
\noindent
($\Leftarrow$)
Now, let us suppose that there is an assignment $\Theta_X$ of truth value
to atoms in $X$ such that for each assignment $\Theta_Y$ of truth values
to atoms in $Y$, the truth value of $\Theta_{X,Y}(\varphi)$ is true.
We define ${U} =\{+val_X(x,\Theta_X(x)),+val_Y(\bot,\bot), assign(\bot,
\bot)\}$. The revised database $\I'$ satisfies $\I'=\I\cup
\{+val_X(x,\Theta_X(x)),+val_Y(\bot,\bot), assign(\bot,\bot)\}$.
It is easy to check that $\<\I',\eta\>$ is consistent (we recall that
$\<\I,\eta\>$ is consistent).

It is clear that the parts (1) and (2) of the request hold after the
repair. We will prove that $satisfied$ holds after the repair.
Let us consider an arbitrary possible world $W$ of $\<\I',\eta\>$. The
integrity constraints imply that for every $y\in Y$ there is exactly one
atom $val_Y(y,\zeta_y)$ and exactly one atom $assign(\zeta_y,v)$, where
$v=true$ or $v=false$. We define $\Theta_Y$ so that for every $y\in Y$,
$\Theta_Y(y)=true$ if $assign(\zeta_y,true)\in W$, and $\Theta_Y(y)=false$,
otherwise. Since $\Theta_{X,Y}(\vph)$ holds, there is a disjunct $d\in D_\vph$
such that $\Theta_{X,Y}(d)$ holds. Thus, $sat(d)$ holds in $LM({P}\cup W)$
and so, $satisfied$ holds in $LM({P}\cup W)$, too.
\hfill$\Box$
\\

%\noindent
%\emph{Corollary~\ref{existRRcorollary}}\\
%Let $\D=\<\I,\IC,{P}\>$, where $\IC$ is a set of integrity constraints,
%and ${P}$ an acyclic Horn program, and let $\RS$ be a request set. The
%problem of deciding whether there is a relevant repair for $(\D,\RS)$ is
%$\Sigma^P_2$-complete.
%
%Proof: There are only finitely many relevant weak repairs. Therefore, a
%$\sqsubseteq$-minimal relevant weak repair exists if and only if a relevant
%weak repair exists. Thus, the result follows from Theorem \ref{existsRR}.
%\hfill$\Box$
%\\
%\begin{proposition}
%\label{checkA-proposition}
%Let $\D=\<\I,\IC,{P}\>$, where $\IC$ is a set of integrity constraints, ${P}$ an acyclic Horn program, $\U$ an update and $\RS$ a request set.
%The problem of deciding whether an update $\U$
%is an arbitrary weak repair for $(\D,\RS)$ is in $\Sigma_2^P$.
%\end{proposition}
%Proof:
%By the definition of an arbitrary weak repair, to decide whether $\U$
%is an arbitrary weak repair we need to guess a non-nullary constant $a$
%and a subset $Q$ of its occurrences and show that:
%\begin{enumerate}
%\item $\U$ is a weak repair for $(\D,\RS)$, and
%\item the update $\U'$ resulting from replacing occurrences of $a$
%in $Q$ is a weak repair for $(\D,\RS)$.
%\end{enumerate}
%Both tasks are in $\Delta_2^P$ (Proposition \ref{checkWR-proposition}) and
%so the assertion follows. \hfill$\Box$

In order to prove Theorem~\ref{existsCR} we need the following proposition.
\noindent
\begin{proposition}~\label{checkCWR-corollary}
Let $\D=\<\I,\IC,{P}\>$, where $\IC$ is a set of integrity constraints,
${P}$ an acyclic Horn program, ${U}$ an update and $\RS$ a request set.
The problem of deciding whether an update ${U}$
is a constrained weak repair for $(\D,\RS)$ is in $\Pi_2^P$.
\end{proposition}
Proof:
By the definition to decide whether ${U}$
is not a constrained weak repair we need to guess a non-nullary constant $a$
and a subset $Q$ of its occurrences and show that:
\begin{enumerate}
\item ${U}$ is a weak repair for $(\D,\RS)$, and
\item the update ${U}'$ resulting from replacing occurrences of $a$
in $Q$ is a weak repair for $(\D,\RS)$.
\end{enumerate}
Both tasks are in $\Delta_2^P$ (Proposition \ref{checkWR-proposition}).
Therefore, the problem of deciding whether ${U}$ is not a
 constrained weak repair is in $\Sigma_2^P$ and the problem of
deciding whether ${U}$ is a
 constrained weak repair is in $\Pi_2^P$.
\hfill$\Box$
\\

\noindent
\textbf{Theorem~\ref{existsCR}.}
\emph{Let $\D=\<\I,\IC,{P}\>$, where $\IC$ is a set of integrity constraints,
and ${P}$ an acyclic Horn program, and let $\RS$ be a request set.
The problems of deciding whether there is a constrained weak repair and
whether there is a constrained repair
for $(\D,\RS)$ are in $\Sigma^P_3$.}% (conjecture: $\Sigma^P_3$-complete).

\noindent Proof:
Firstly observe that there are only finitely many constrained weak repairs. Therefore, a
$\sqsubseteq$-minimal constrained weak repair exists if and only if a constrained
weak repair exists. In the following we prove that the problem of deciding whether there is a constrained weak repair
for $(\D,\RS)$ is $\Sigma^P_3$-complete.

Each constrained weak repair is a relevant weak repair. Therefore,
its size is polynomial in the size of $\D$. Once we guess a relevant update
${U}$, we have to test if it is a constrained weak repair. We know, by Proposition
\ref{checkCWR-corollary}, that the second problem is in $\Pi_2^P$. Therefore,
the problem in the class $NP^{\Pi_2^P}=\Sigma^P_3$.
\hfill$\Box$
\\

%\noindent
%\emph{Corollary~\ref{existCRcorollary}}\\
%Let $\D=\<\I,\IC,{P}\>$, where $\IC$ is a set of integrity constraints,
%and ${P}$ an acyclic Horn program, and let $\RS$ be a request set.
%The problem of deciding whether there is a constrained repair
%for $(\D,\RS)$ is in $\Sigma^P_3$.% (conjecture: $\Sigma^P_3$-complete).

\end{document}